\documentclass[fleqn,usenatbib]{mnras}

\usepackage[utf8]{inputenc}
\usepackage{ae,aecompl}

\usepackage{graphicx}	
\usepackage{amsmath}	
\usepackage{amssymb}	
\usepackage{soul}
\usepackage[normalem]{ulem}
\usepackage[dvipsnames]{xcolor}
\usepackage[normalem]{ulem}
\usepackage{cancel}
\usepackage{lineno}
\usepackage{float}

\linenumbers

\newcommand{\Msun}{\,M$_\odot$}

\usepackage{newtxtext,newtxmath}

\title[ICL in galaxy proto-clusters]{Intracluster light in the core of z$\boldsymbol{\sim}$2 galaxy proto-clusters}

\author[S. V. Werner et al.]{
S. V. Werner$^{1}$\thanks{E-mail: stephanevazwerner@gmail.com},
N.\,A.~Hatch$^{1}$,  J. Matharu$^{2,3,4}$, A. H. Gonzalez$^{5}$,  \newauthor
Y. M. Bahé$^{6}$, S. Mei$^{7,8}$, G. Noirot$^{9}$ \& D. Wylezalek$^{10}$\\
\\
$^{1}$School of Physics and Astronomy, University of Nottingham, Nottingham, NG7 2RD, UK\\
$^{2}$Department of Physics and Astronomy, Texas A\&M University; College Station, TX, 77843-4242, USA\\
$^{3}$George P. and Cynthia Woods Mitchell Institute for Fundamental Physics and Astronomy, Texas A\&M University; College Station, TX, 77845-4242, USA \\
$^{4}$Cosmic Dawn Center, Niels Bohr Institute, University of Copenhagen, Rådmandsgade 62, 2200 Copenhagen, Denmark\\
$^{5}$Department of Astronomy, University of Florida; 211 Bryant Space Science Center, Gainesville, FL 32611, USA\\
$^{6}$Leiden Observatory, Leiden University; Niels Bohrweg 2, NL-2333 CA, Leiden, The Netherlands \\
$^{7}$Université de Paris, CNRS, Astroparticule et Cosmologie; F-75013 Paris, France \\
$^{8}$Jet Propulsion Laboratory and Cahill Center for Astronomy \& Astrophysics, California Institute of Technology; 4800 Oak Grove Drive, Pasadena,\\
California 91011, USA\\
$^{9}$Department of Astronomy \& Physics, Saint Mary’s University; 923 Robie Street, Halifax, NS B3H3C3, Canada\\
$^{10}$Zentrum für Astronomie der Universität Heidelberg, Astronomisches Rechen-Institut, Mönchhofstr 12-14, D-69120 Heidelberg, Germany  \\
}

\date{Accepted XXX. Received YYY; in original form ZZZ}

\pubyear{2023}

\begin{document}
\nolinenumbers
\label{firstpage}
\pagerange{\pageref{firstpage}--\pageref{lastpage}}

\maketitle

\begin{abstract}
Intracluster light is thought to originate from stars that were ripped away from their parent galaxies by gravitational tides and galaxy interactions during the build up of the cluster. The stars from such interactions will accumulate over time, so semi-analytic models suggest that the abundance of intracluster stars is negligible in young proto-clusters at $z\sim2$ and grows to around a quarter of the stellar mass in the oldest, most mature clusters. In contrast to these theoretical expectations, we report on the detection of intracluster light within two proto-clusters at $z=2$ using deep {\it HST} images. We use the colour of the intracluster light to estimate its mass-to-light ratio in annuli around the brightest cluster galaxies (BCG), up to a radius of 100\,kpc. We find that $54\pm5$\% and $71\pm3$\% of the stellar mass in these regions is located more than 10\,kpc away from the BCGs in the two proto-clusters. This low concentration is similar to BCGs in lower redshift clusters, and distinct from other massive proto-cluster galaxies.  This suggests that intracluster stars are already present within the core 100\,kpc of proto-clusters.
We compare these observations to the Hydrangea hydrodynamical galaxy cluster simulations and find that intracluster stars are predicted to be a generic feature of group-sized halos at $z=2$. These intracluster stars will gradually move further away from the BCG as the proto-cluster assembles into a cluster.

\end{abstract}




\section{Introduction}

In the standard cosmological paradigm small density fluctuations of dark matter in the early Universe rapidly collapsed into triaxial structures called halos, which provided a gravitational well deep enough to trap gas and produce the first stars. Over the following 13 billion years, these halos merge to produce progressively larger halos. What happens to the stars in these merging halos is a matter of debate that is pivotal to our understanding of the evolution of galaxies. 

Clusters of galaxies are the most massive halos in the Universe and are therefore the most extreme examples of hierarchical merging. As such, their central galaxies, known as brightest cluster galaxies (BCGs), are uniquely suited to study the process of hierarchical galaxy formation. Early galaxy formation models \citep{Delucia_2007} predicted that BCGs would undergo protracted growth, in step with the growth of their dark matter halos. But observations of distant BCGs conflicted with these predictions and demanded only modest growth in both size and mass since z$\sim$1 \citep{Collins_2009, Whiley_2008, Chu_2021}. To account for this lack of growth, some models were updated to remove a fraction of the stars from the central galaxies of the merging halos and deposit them as free-floating stars in the merged halo which is visible as diffuse, intracluster light \citep{Contini_2014}. These new models produce modest BCG growth since z$\sim$1 and a corresponding rapid increase in intracluster light over the same period. As a result, these models predict negligible amounts of intracluster light in z$\sim$2 proto-clusters \citep{Contini_2021}. On the other hand, intracluster stars have been observed in massive clusters up to $z=1.75$ \citep{Demaio_2020} {  and $z=1.85$ \citep{Joo_Lee_2023}}, and diffuse UV light has been observed in a proto-cluster at $z=2.2$ \citep{Hatch_2008}. Therefore, intracluster light appears to be a generic feature of halo assembly, even early on, and not just a low-redshift phenomenon. Recent reviews of intracluster light observations and theory can be found in \citet{Montes_2022} and \citet{Contini_2021_review}.

To test these models we investigate the amount of intracluster light present within the core of two galaxy proto-clusters at z$\sim$2. CARLA\,J$1018+053$ (hereafter CARLA\,J1018) was discovered to be an over-density of galaxies around a radio-loud quasar at z$\sim$1.95 \citep{Noirot_2018, Wylezalek_2013}, with potential intracluster light associated with the over-density \citep{Noirot_2018}. XLSSC-122 was first identified as a faint X-ray source by the XMM Large Scale Structure survey \citep{Willis_2013}, which was subsequently discovered to belong to a $M_{500} = 6.3\pm1.5 \times 10^{13}$ \Msun\  halo \citep{Mantz_2018} and spectroscopically confirmed to be a large galaxy over-density at $z=1.98$ using  {\it HST} grism spectra \citep{Willis_2020, Noordeh_2021}.
Both proto-clusters contain similar amounts of stars ($\sim10^{12}$ \Msun), indicating that the two regions may contain a similar amount of dark matter. However, XLSSC-122 has a large mass gap between its first and second most massive galaxies, unlike CARLA\,J1018, whose three most massive galaxies have comparable masses. This suggests that XLSSC-122 has already assembled much of its dark matter into a common halo, while CARLA\,J1018 is still a sprawling proto-cluster consisting of several lower-mass halos \citep{Golden_Marx_2022}. 

The separation between BCG and intracluster light is difficult to define observationally. Some authors use a surface brightness limit to separate between the two \citep[e.g.][]{Burke_2015}, whereas others fit multiple S\'ersic profiles to the light and define the intracluster light to be the outermost, flat component \citep[e.g.][]{Zhang2019,Joo_Lee_2023}. In this work we do not try to separate the two structures due to the technical challenge of observing the faint intracluster light at high redshifts. Instead, we infer the presence of intracluster light in systems which have low concentrations of stars within the central 100\,kpc of the proto-cluster. Measuring the light concentration within this radial distance is justified by the recent measurements of \citet{Joo_Lee_2023} who showed that the BCG typically dominates only the central 10\,kpc region (or $\sim$30\,kpc if the BCG is fit with two S\'ersic components) for clusters above $z=1.3$. Therefore, we will infer the presence of intracluster light in proto-clusters which exhibit low stellar mass and light concentrations within the central 10\,kpc compared to the area encompassed by a radius of 100\,kpc around the BCGs, and if they contain significant light at a radial distance beyond 50\,kpc from the BCG core.

Throughout this paper, we assume a flat, $\Lambda-$cold dark matter cosmological model, parameterized by $\mathrm{\Omega_{M}}$=0.315, $\mathrm{\Omega_{\Lambda}}$=0.685,$\mathrm{H_{0}}$=67.4 km $\mathrm{s^{-1}Mpc^{-1}}$ \citep{Planck_2020}, and all magnitudes are based on the AB system.

\section{Data}
\subsection{Protocluster sample}

The protoclusters for this study were selected from a sample of four z$\sim$2 spectroscopically confirmed proto-clusters: XLSSC-122 at $z = 1.98$ \citep{Willis_2020}, CARLA\,J$1018+053$ at $z = 1.96$, CARLA\,J$0800+4029$ and CARLA\,J$2039-2514$ both at $z = 2.00$  \citep{Wylezalek_2014, Noirot_2018}. CARLA\,J$0800+4029$ and CARLA\,J$2039-2514$ were deemed unsuitable for this study due to contamination of the BCG and the surrounding intracluster light: a large foreground galaxy and bright star lies along the line of sight to the BCG of CARLA\,J$2039-2514$, and the most massive galaxy in CARLA\,J$0800+4029$ is likely to be the luminous $[F140W]=19.7$ radio-loud quasar, SDSS\,J$0800+4029$. 

\subsection{Hubble Space Telescope Data}

High resolution images for XLSSC 122 were harvested from the {\it HST} archive (programme ID 15267). The WFC3 images were taken through the $F140W$ filter at four different orientations (across 12 orbits), with a total exposure time of 5,171 seconds, and one orbit of data was taken through the $F105W$ filter, observed at a single orientation, that resulted in a total exposure time of 2,611 seconds. High resolution images and grism spectra were harvested for CARLA\,J1018 from programme ID 13740. The $F140W$ images were taken in two orbits, each at a different orientation, for a total of 973 seconds. The remaining time of each orbit was used to observe through the G141 grism, details of which are provided in \citet{Noirot_2016, Noirot_2018}.

We performed full end-to-end processing of the {\it HST} images and grism data using the grism redshift \& line analysis software for space-based slitless spectroscopy \citep{Brammer_2019}. The images were flat-fielded using both pixel to pixel flat fields and a delta correction. The sky background is fit during the cosmic ray correction step {\it CRCORR} by computing a linear fit to the accumulated signal from each readout. The astrometric registration uses {\it GAIA eDR3} data to perform a fine alignment and a final combined mosaic of all orientations for each filter was achieved using Astrodrizzle and the default distortion correction tables. The output images have a pixel scale of 0.06 arcsec. The {\it HST} images were masked to remove bad pixels and pixels with less than 66\% of the median image weight (measured from the drizzled weight images). Furthermore, the $F105W$ image of XLSSC 122 exhibited a defect which was masked using a rectangular aperture of 10.8 by 40\,arcsec, centred on [$34.4286$, $-3.7694$] and angled by $20^{\circ}$.

We then extracted the grism spectra from the CARLA\,J1018 data. A source model was constructed from the science image, which was then used to produce a contamination model for the G141 grism spectra. The contamination model was subtracted and finally the clean G141 grism spectra were extracted.

\subsection{Ground-Based Images}

Images of CARLA\,J1018 at broad wavebands of $z$, J, H, Ks and two narrow-bands at 1.06 and 1.19µm were obtained using {\it FORS} and {\it HAWK-I} instruments on the Very Large Telescope, {\it ESO} via programmes 094.A-0343 and 096.A-0317. The 1.06 and 1.19µm narrow-band images were obtained because they tightly bracket the 4000Å and Balmer breaks of galaxies at z$\sim$2, so greatly improve the photometric redshift, age and mass measurements of the proto-cluster galaxies. The near-infrared {\it HAWK-I} data were reduced using standard near-infrared reduction techniques with the {\it ESO MVM} software \citep{Vandame_2004}. The {\it FORS} $z-$band data were reduced using the theli data reduction pipeline \citep{Schirmer_2013}. We added an archival $i-$band image of this field taken on the William Herschel Telescope \citep{Cooke_2015}.

Since the $z-$band image is used to measure the colour of the intracluster light, we ensured the sky background is as flat as possible through the following processes. Each exposure was first flat-fielded using dome flats. Then a single static background model was created from all the exposures taken in each night. The sources were first masked, then exposures were median combined without any astrometric correction (to correct for the dithering) to produce a sky background model. The structure in the model included low-level fringing and gradients. Each exposure was then corrected for this background model. Finally, we eye-balled each exposure for satellite trails, anomalous reflected light, and poor chip regions and masked these regions on the individual exposures before they were coadded together to produce the final science-grade image.

Flux and astrometric calibration for J, H and Ks images was achieved using {\it 2MASS} catalogues \citep{2mass_2006}. For the other ground-based images, relative flux calibration is done based on the universal properties of the stellar locus using stellar libraries \citep{Pickles_1998, Ivanov_2004} and applying offsets to the instrumental magnitudes so that colours of stars in the images match the reference locus. Finally, we applied Galactic extinction corrections \citep{Schlegel1998}. Properties of these ground-based data are presented in Table \ref{tab:ground}.

\begin{table}
\begin{center}
\caption{Properties of the ground-based images of CARLA\,J1018 used to construct the proto-cluster galaxy catalogue in Table \ref{tab:members}. Image depths were measured in 2 arcsec diameter apertures.}
\begin{tabular}{l c c}

\hline
Filter & 3$\sigma$ image depth & PSF FWHM (arcsec) \\
\hline

$i$ & 25.59 & 0.80\\
$z$ & 25.14 & 0.64 \\
NB 1.06 & 23.69 & 0.75 \\
NB 1.19 & 23.68 & 0.60 \\
J & 24.58 & 0.64 \\
H & 23.47 & 0.44 \\
K & 23.48 & 0.64 \\

\hline
\end{tabular}
\label{tab:ground}
\end{center}
\end{table}

The J, H and Ks images were combined to create a deep image from which we detected sources using SExtractor \citep{Bertin_1996}. Since the resolution of the images varied, we Gaussian-smoothed each to match the image with the poorest resolution (the $i-$band image) before fluxes were measured within 2-arcsec-diameter circular apertures. Total fluxes were obtained by applying an aperture correction determined from the growth curves of unsaturated stars. Uncertainties on the fluxes were taken to be the square root of the photon counts in the apertures plus the standard deviation of the total photon counts within 2-arcsec-diameter apertures placed in empty regions of the images. The final object catalogue was cleaned by removing any source that was located within the regions with less than 30\% of the total observing time in each of the seven ground-based images.

\begin{table}
 \caption{Surface brightness limits. Limiting surface brightness limits for CARLA\,J1018 and XLSSC-122 images defined as the 3$\sigma$ limit derived over a 100 $\mathrm{arcsec^{2}}$ area.}
\centering
 \label{tab:sb}
 \begin{tabular}{lcc}
  \hline
Cluster     & Filter & $\mathrm{\mu_{lim} (mag/ arcsec^{2})}$ \\ \\
  \hline
 \centering
CARLA\,J1018 & $F140W$  & 29.2             \\
CARLA\,J1018 & z      & 27.1             \\
XLSSC-122   & $F140W$  & 30.3             \\
XLSSC-122   & $F105W$  & 30.3    \\

\hline
 \end{tabular}
\end{table}

\subsection{Sky subtraction and surface brightness limits}
\label{sec:skysub}
Intracluster light has a low surface brightness and therefore is particularly sensitive to errors in the measurement and subtraction of the background light. It is therefore important to robustly measure the sky background and estimate the uncertainty in the four images we use to calculate the luminosity and colour of the intracluster light. The {\it HST} images will not be affected by atmospheric emission, and neither target is in a region contaminated by significant amounts of Galactic cirrus, so we expect the background light of these images to be dominated by Zodiacal light and exhibit a smooth distribution over the whole image. However, the $z-$band image comes from a ground-based instrument and is therefore subject to a variable atmospheric sky background, whilst the $F105W$ can be affected by a time varying background due to 1.083 $\mu$m He I emission line from the Earth’s atmosphere.

We first measure and subtract a global background from the images. To do this, we detected all sources using SExtractor. We used the default {\sc sextractor} parameters but change the following: {\sc detect\_minarea}=5, {\sc detec\_thresh}=1.9, {\sc analysis\_threshold}=1.9, {\sc deblend\_mincont}=0.005 for CARLA\,J1018, and further updated {\sc detect\_thresh}=1.7, and {\sc analysis\_threshold} = 1.7 to detect the fainter sources in the deeper XLSSC image. We then masked all sources to 8 times their semi-major and minor axes to ensure no source light was visible. Finally, we masked the entire 150 kpc region around the BCG to ensure our background measurement was not affected by any potential intracluster light.

We fit a Gaussian to the pixel flux distribution from these masked images. The  centre of the peak of the Gaussian fit is taken as the global background and subtracted from the original images.  By subtracting this background pedestal we also remove any diffuse light from unresolved faint galaxies not associated with the proto-clusters. The standard deviation of the pixel flux distribution, $\sigma$, is used to derive the limiting magnitudes through $\mathrm{\mu_{lim} = Z_{p} - 2.5 \times\log \frac{3\sigma}{pix \sqrt{\Omega}}}$, where $\mathrm{Z_{p}}$ is the zero point of the image, pix is the pixel scale and $\Omega$ is the solid angle in $\mathrm{arcsec^{2}}$ over which the limiting magnitude is defined. This equation only remains valid over small solid angles where the noise is dominated by pixel-to-pixel variations and Poisson noise. In order to readily compare with the depths of current and future surveys, e.g., \citet{euclid2_2022}, we list the limiting $3\sigma$ surface brightness of these images in Table \ref{tab:sb} calculated over 100\,$\mathrm{arcsec^{2}}$.

We quantify whether any large-scale variations in the background are present in the four images we use to measure the intracluster light flux and colour. To do this we placed 5.8 arcsec-radius apertures (corresponding to 50 kpc at the redshift of the proto-clusters) in random locations over the masked images. The distribution of this background measurement conforms to a Gaussian for the {\it HST} images, which implies that the background noise is uncorrelated and does not contain large variations on this scale. The background in the ground-based z-band image is well matched to a Gaussian but has a tail to low fluxes. Only 6\% of the apertures had anomalously low fluxes. An investigation into the $z-$band image showed that these low fluxes were caused by a clustering of individual low-value pixels in a certain region of the image (far from the BCG of the proto-cluster). Thus, there is no evidence that there are large-scale variations in the background of the $z-$band image. Clustered low-valued pixels were not observed near the BCG and so this should not affect the measured colour of the intracluster light.


\section{Methodology}

\setlength{\tabcolsep}{2pt}

\begin{table}
 \caption{Properties of 15 members of CARLA\,J1018 identified through template fits to the combined grism spectroscopy and photometry. A map of their locations in the proto-cluster is shown in Figure \ref{RA_DEC}.}
\centering
 \label{tab:members}
 \begin{tabular}{lccccc}
  \hline
\centering
Label  & Position  & Log($\mathrm{M_{*}/M_{\odot}}$) & $\mathrm{I_{mem}}$ & mag & Observed \\

 & (J2000) & & & &  M/L \\
 \\
  \hline
Quasar & \begin{tabular}[c]{@{}l@{}}10h18m27.8s \\                        +05d30m29.9s\end{tabular}  & -                                        & 1.0  & 19.47 & -            \\
{\it a}      & \begin{tabular}[c]{@{}l@{}}10h18m30.5s \\                        +05d31m00.6s\end{tabular}  & 11.59                                    & 0.83 & 21.50 & 1.73         \\
{\it b}      & \begin{tabular}[c]{@{}l@{}}10h18m29.9s \\                        +05d31m04.6s\end{tabular}  & 11.41                                    & 0.98 & 21.27 & 0.93         \\
{\it c}      & \begin{tabular}[c]{@{}l@{}}10h18m30.0s \\                         +05d31m06.6s\end{tabular} & 11.24                                    & 0.53 & 21.48 & 0.76         \\
{\it d}      & \begin{tabular}[c]{@{}l@{}}10h18m30.7s \\                         +05d30m55.0s\end{tabular} & 11.08                                    & 0.95 & 21.49 & 0.53         \\
{\it e}      & \begin{tabular}[c]{@{}l@{}}10h18m30.0s \\                         +05d31m04.1s\end{tabular} & 11.03                                    & 0.37 & 21.50 & 0.48         \\
{\it f}      & \begin{tabular}[c]{@{}l@{}}10h18m28.6s \\                        +05d29m43.9s\end{tabular}  & 11.00                                    & 0.79 & 22.09 & 0.77         \\
{\it g}      & \begin{tabular}[c]{@{}l@{}}10h18m30.1s \\                         +05d31m00.6s\end{tabular} & 10.77                                    & 0.40 & 23.27 & 1.34         \\
{\it h}      & \begin{tabular}[c]{@{}l@{}}10h18m30.6s \\                         +05d30m56.6s\end{tabular} & 10.71                                    & 0.16 & 23.14 & 1.04         \\
{\it i}      & \begin{tabular}[c]{@{}l@{}}10h18m30.7s \\                        +05d30m57.2s\end{tabular}  & 10.57                                    & 0.10 & 22.55 & 0.44         \\
{\it j}      & \begin{tabular}[c]{@{}l@{}}10h18m28.3s \\                         +05d29m29.8s\end{tabular} & 9.76                                     & 0.95 & 23.19 & 0.12         \\
{\it \#162}   & \begin{tabular}[c]{@{}l@{}}10h18m28.1 \\                         +05d29m30.68s\end{tabular} & -                                        & 1.0  & 24.20 & --           \\
{\it \#446}   & \begin{tabular}[c]{@{}l@{}}10h18m28.2   			\\  			 +05d30m19.71s\end{tabular}               & -                                        & 0.55 & 24.23 & --           \\
{\it \#354}   & \begin{tabular}[c]{@{}l@{}}10h18m27.4 \\                       +05d30m34.47s\end{tabular}   & -                                        & 0.99 & 24.63 & --           \\
{\it \#336}   & \begin{tabular}[c]{@{}l@{}}10h18m31.2s \\                       +05d31m13.2s\end{tabular}   & \multicolumn{1}{l}{--}                   & 1.0  & 24.11 & --          \\

\hline

\end{tabular}
\end{table}

\begin{figure*}
\includegraphics[width=1.5\columnwidth]{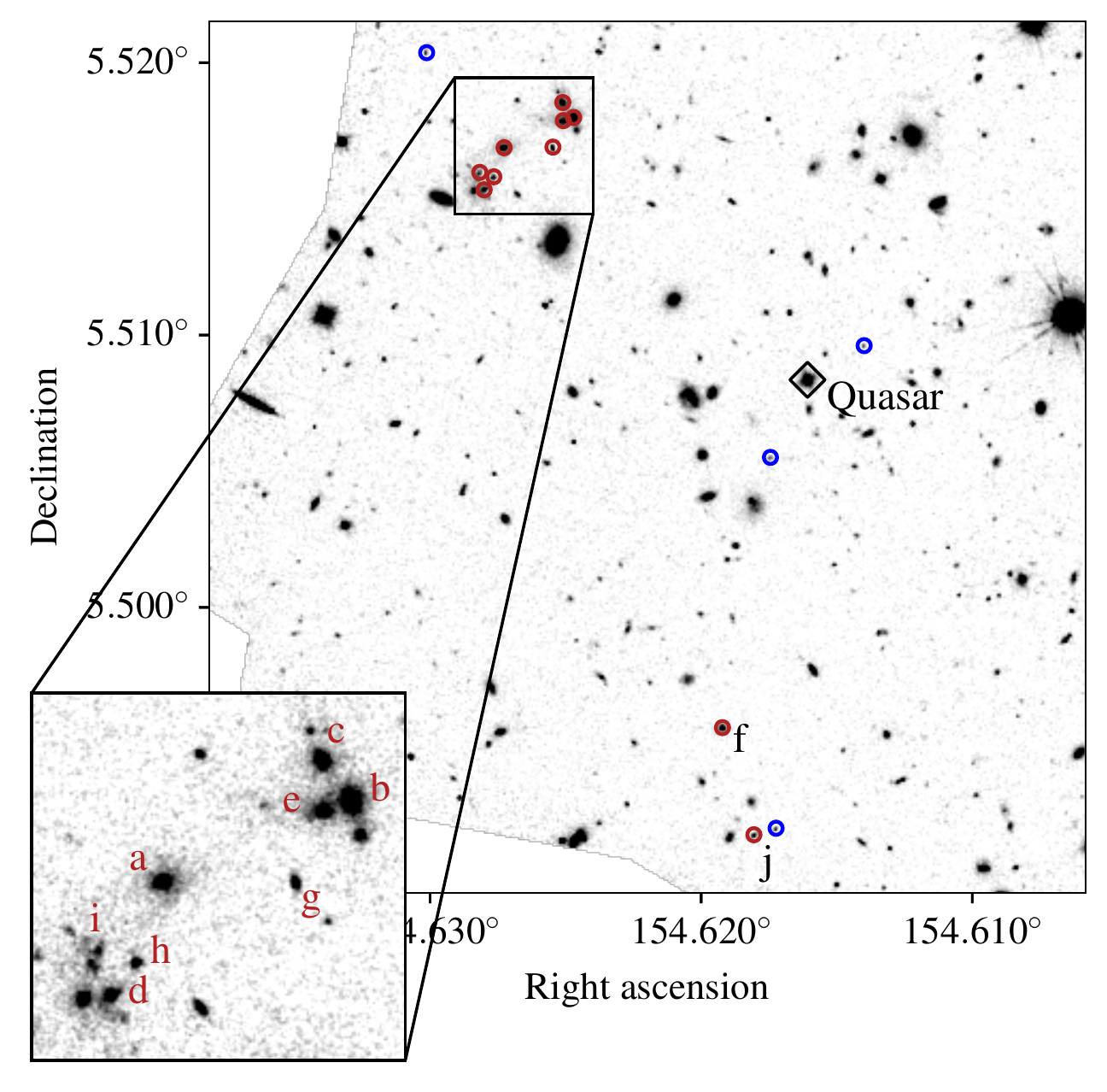}
\caption[parameters]{The {\it HST} $F140W$ image of CARLA\,J1018 with marked proto-cluster galaxies. Red circles mark proto-cluster galaxies with strong continuum emission; blue circles mark emission-line galaxies that do not have enough continuum to allow us to measure their mass. The radio-loud quasar is marked by a black diamond. The insert details the region with the highest galaxy density that we take to be the proto-cluster core, with galaxy a chosen as the BCG.}
\label{RA_DEC}
\end{figure*}

\subsection{Identifying the BCG of CARLA\texorpdfstring{$\,$}{}J1018}

\begin{figure*}
\includegraphics[width=1.9\columnwidth]{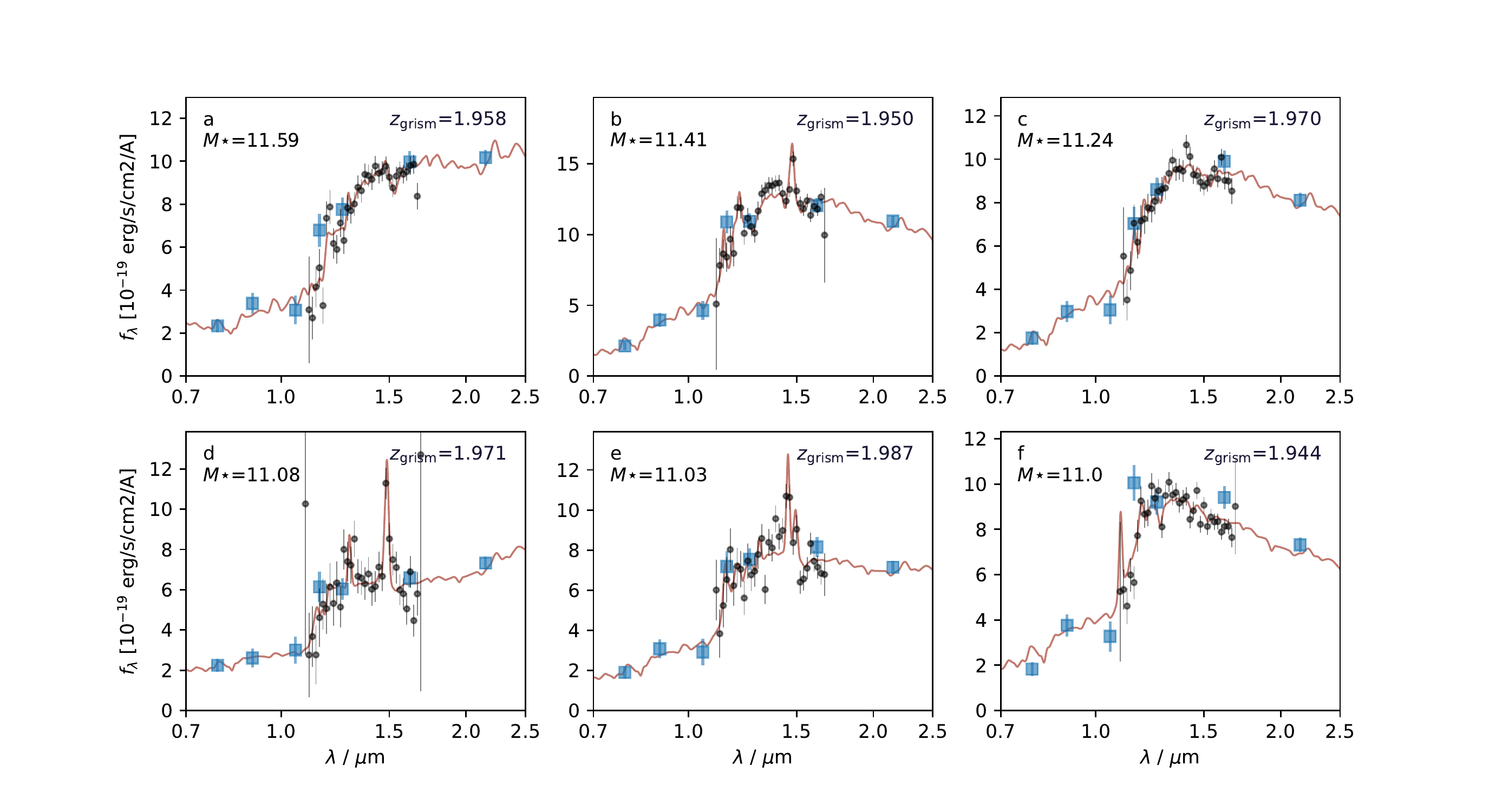}
\caption[parameters]{Spectral energy distributions of the six most massive galaxies in CARLA\,J1018. The black circles show the grism spectral data while the blue squares show the photometry from the ground-based images. The galaxy template that best fits the spectral energy distribution is shown in red.}
\label{spectra}
\end{figure*}

To identify the BCG of CARLA\,J1018 we first construct a galaxy member catalogue using both photometric and grism data.
We used a two-step process to identify the members of the CARLA\,J1018 proto-cluster within the {\it HST} field-of-view. We first fitted the ground-based photometry with a suite of galaxy templates using {\sc eazy} version 0.5.2 \citep{Brammer_2008} using no redshift prior. The templates were stepped in redshift from 0.01 to 6 over a grid of $\Delta z/(1+z)=0.01$. During this fitting we iteratively adjusted the zeropoints of the photometry to minimize the template fit residuals. We then combined the zeropoint-adjusted photometric data and grism spectroscopy for the 169 sources that were detected in both data sets. 

We then fitted these data with a suite of galaxy templates known as Flexible Stellar Population Synthesis models \citep{Conroy_2009, Conroy_2010} using the software {\sc grizli} version 1.3.2 \citep{Brammer_2019}. The templates were stepped in redshift over a coarse grid of $\Delta z=0.004$ between the 68\% confidence redshift intervals determined by the {\sc eazy} fitting procedure. Finally, the templates were stepped in redshift over a fine grid of $\Delta z=0.001$ around the peaks in the redshift probability distribution.

There are 275 sources which are detected in the {\it HST} image that are too faint to have counterparts in the ground-based images. These sources have low levels of continua but can be strong line emitters, so we fit their grism spectra with galaxy templates stepped in redshift from 0.01 to 6 over the same coarse and fine grids as before. We visually inspected each of the fits and removed any spectral extractions that were unreliable due to poor modelling of the contamination by nearby sources.

The best-fit redshifts from the template fitting had uncertainties of typically $\Delta z=0.006$ for emission line galaxies with strong continuum (e.g. galaxy {\it j} in Figure \ref{RA_DEC}) or with extremely strong continuum and strong Balmer/4000\AA \ breaks (e.g. galaxy {\it a}). However, for galaxies without strong emission lines the typical uncertainty was much larger at $\Delta z=0.05$. Therefore, selecting protocluster members by their best-fit redshifts could introduce a bias against locating passive proto-cluster galaxies. We therefore identified proto-cluster members using two criteria.

We first selected galaxies with a best-fit redshift within a broad window of $z\mathrm{_{cl} \pm 0.05}$, then removed those galaxies that did not have a highly peaked redshift probability distribution function over the redshift interval $z\mathrm{_{cl} \pm 0.02}$. This interval corresponds to $\pm$2000 km/s, which is three times the typical velocity dispersion of massive clusters at this redshift \citep{Willis_2020}. We defined  $\mathrm{I_{mem}}$ as the integral of the redshift probability distribution function over the redshift interval $z\mathrm{_{cl} \pm 0.02}$, and calculated this for each galaxy. We start with an initial guess of the cluster's redshift at $z\mathrm{_{cl}}=1.95$, determined by previous work \citep{Noirot_2018}, then iteratively redefine the cluster's redshift as the mean redshift of galaxies with $\mathrm{I_{mem}}>0.5$ resulting in  $z\mathrm{_{cl}} =1.96$. $\mathrm{I_{mem}}$ is strongly dependent on the signal-to-noise of the data so galaxies with strong line emission or continuum have larger $\mathrm{I_{mem}}$ than lower luminosity members. We therefore eye-balled the redshift probability distribution functions and settled on the choice of $\mathrm{I_{mem}}> 0.1$ as defining the redshift probability distribution function as being ‘highly peaked.’ This value is low enough that it did not rule out low luminosity galaxies with Balmer/4000\AA \ breaks that were distinguishable by eye (such as galaxy {\it i}), but high enough that it removed galaxies without any clear features in the observed spectral energy distribution or the redshift probability distribution function. Using the criteria of $\mathrm{I_{mem}}>0.1$ and having a best-fit redshift within $z\mathrm{_{cl} \pm 0.05}$ resulting in 15 proto-cluster members, which we show in Figure \ref{RA_DEC} and list in Table \ref{tab:members}. The photometry and best-fit stellar population templates for the six most massive proto-cluster members are provided in Figure \ref{spectra}.

To obtain stellar masses we fit the ground-based photometry with stellar population templates using FAST \citep{Kriek_2009}. The stellar population templates are based on Bruzual \& Charlot models \citep{BC_03} with exponentially declining star formation histories, dust attenuation and a Chabrier initial mass function \citep{Chabrier_2003}. The redshift of the cluster members is fixed at $z\mathrm{_{cl}} =1.96$, but the masses do not change significantly if the redshift is fixed to the most probable grism redshift. Table \ref{tab:members} lists the properties of the 15 proto-cluster members, including the radio-loud quasar SDSSJ101827+0530 that was first used to identify this cluster. We estimate the observed $F140W$ mass-to-light for each galaxy using their stellar masses and observed-frame $F140W$ luminosities.

We identify 6 common members with the original membership catalogue from \cite{Noirot_2018}: the quasar, galaxy {\it j} (labelled {\it \#138} in \citealt{Noirot_2018}), and all numbered galaxies in Table \ref{tab:members}. All numbered galaxies copy the labels given in \citet{Noirot_2018}. The addition of the ground-based photometry allowed us to reclassify two possible members selected by \citet{Noirot_2018}, their {\it \#127} and {\it \#647}, as being H$\alpha$ emitters at a lower redshift. Galaxies labelled {\it a} – {\it i} are new detections enabled by the additional photometry and the grizli software used to reanalyse the grism data. These new detections have weaker emission lines but stronger Balmer and 4000\AA \ breaks than the galaxies in the original catalogue.

The selection of a BCG for the proto-cluster from Table \ref{tab:members} is not straight-forward as the galaxies labelled {\it a} to {\it d}, as well as the quasar, could all be contenders for the most massive galaxy in the proto-cluster. To differentiate between these galaxies, we consider each galaxy’s local density because the barycentre of the proto-cluster is likely to be the region with the highest galaxy density. We rule out the quasar based on its relative isolation compared to the high-density region around galaxies {\it a} – {\it d} shown within the insert of Figure \ref{RA_DEC}. Out of galaxies {\it a} – {\it d}, we select galaxy {\it a} as the BCG because it is the most massive and has the highest galaxy density on the scale of 100 kpc out of all galaxies in the proto-cluster.

\subsection{Identifying the BCG of XLSSC-122}

We use the XLSSC-122 proto-cluster membership catalogue presented in  \cite{Noordeh_2021} and stellar masses of the XLSSC-122 galaxies that were derived from 2-band photometry in \cite{Willis_2020}. However, to compare them with the stellar masses we derive for the CARLA\,J1018 galaxies, we convert the stellar masses from a Salpeter IMF to Chabrier IMF by multiply the masses by 0.61. From this galaxy catalogue, we select the most massive galaxy as the BCG, which is galaxy {\it \#529} at $\mathrm{RA}= 34.4342$, $\mathrm{Dec}= -3.7588$, with $[F140W] = 20.64$ and a stellar mass of $5\times10^{11}$\Msun. This galaxy lies near the centre of the X-ray contours \citep{Willis_2020}, and therefore is likely to lie close to the centre of the gravitational well of the most massive dark matter halo in the proto-cluster. Due to the longer wavelength coverage of the data on the CARLA\,J1018 field, the masses of the CARLA\,J1018 galaxies are likely to be more accurate compared to those of the XLSSC 122 catalogue, which were derived from only 2-band photometry ($F105W$ and $F140W$).

\section{Results}

\begin{figure*}
\includegraphics[width=1.5\columnwidth]{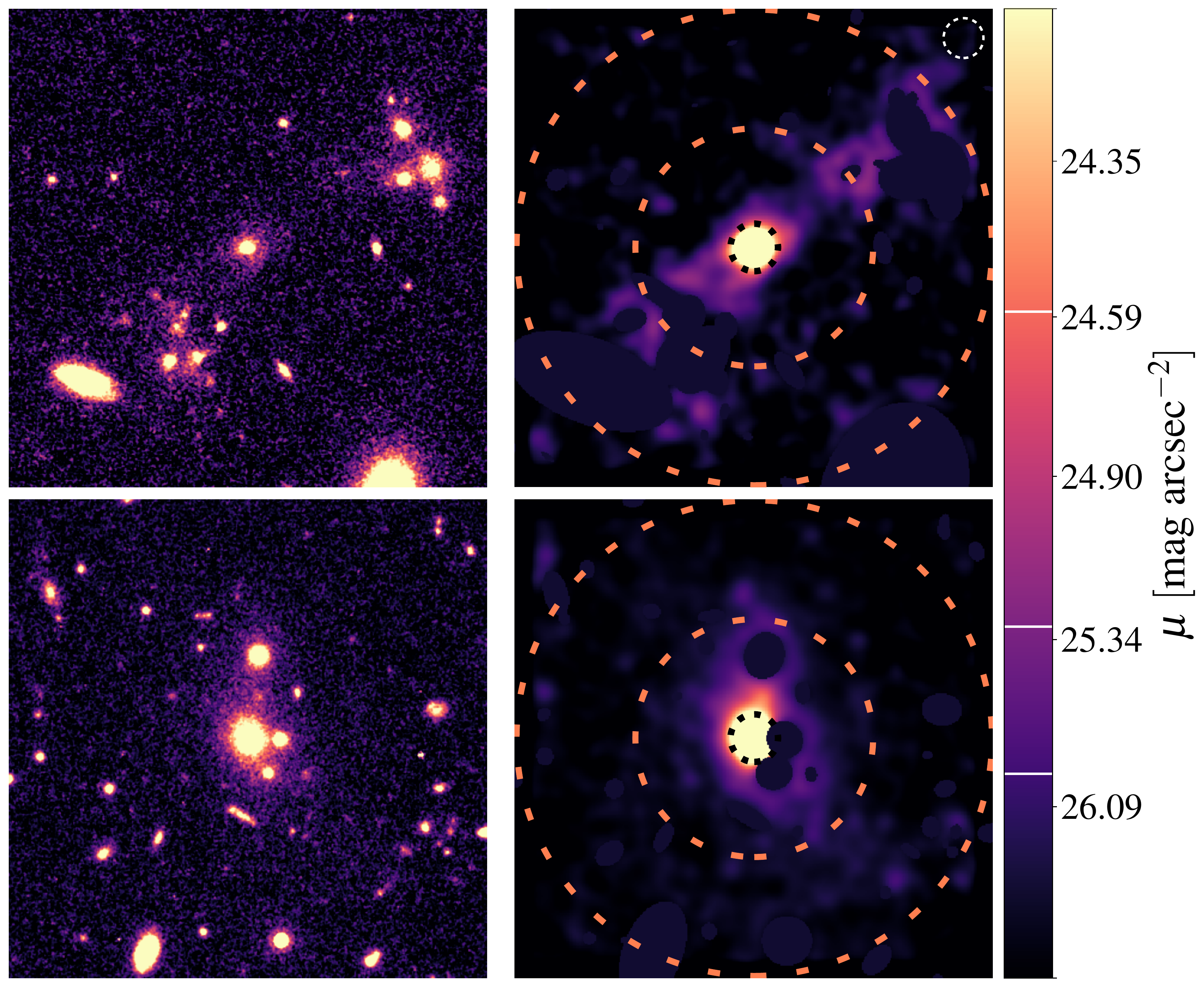}
\caption[parameters]{{\it HST} images of the core of the CARLA\,J1018+0530 (upper panels) and XLSSC 122 (lower panels) proto-clusters (F140W band). The left images show the BCG, intracluster light and surrounding galaxies at the native resolution of the {\it HST} images.  On the right, we highlight the light from the intracluster stars by masking all galaxies, except for the BCG, then smoothing the remaining pixels with a Gaussian 2D kernel of $\sigma =0.12$\,arcsec. The dashed circles mark a radial distance from the BCG centre of 10, 50 and 100\,kpc. {  The white lines on the colour bar marks the 3$\sigma$ (7$\sigma$), 5$\sigma$ (12$\sigma$), and 10$\sigma$ (24$\sigma$) depths in the CARLA\,J1018+0530 (XLSSC 122) images within an area of 1 sq arcsec, as shown by the white circle in the top right.}}
\label{ICL}
\end{figure*}
 
 \setlength{\tabcolsep}{7pt}

\begin{table*}
\centering
 \caption{The $F140W$ surface brightness, $F140W$ luminosity and stellar mass of the BCG \& intracluster light within the z$\sim$2 proto-clusters. The luminosity was measured inside circular apertures and annuli centered on the BCG. The uncertainties account for any residual variations in the background as they are measured from the standard deviation of flux measured in $\sim$1000 randomly distributed apertures across each image. Note that the stellar mass uncertainties are purely statistical and do not account for systematic uncertainties due to the uncertain mass-to-light ratio {  (see Section \ref{sec:mass2light} for details). The average surface brightness within each aperture is greater than the $3\sigma$ surface brightness limit of the images that is listed in Table \ref{tab:sb}. } \label{ICL_lum}}
 \label{tab:lum}
 \begin{tabular}{c|ccc|ccc|ccc|}
  \hline
    &   \multicolumn{3}{c}{Surface brightness (mag/arcsec$^2$)}    &   \multicolumn{3}{c}{Luminosity ($\mathrm{10^{11} L_{\odot}}$)} & \multicolumn{3}{c}{Stellar mass ($\mathrm{10^{11} M_{\odot}}$)}\\
  \hline
Aperture (kpc) & <10 & 10 – 50 & 50 – 100 & <10 & 10 – 50 & 50 – 100 & <10 & 10 – 50 & 50 – 100\\

\hline

CARLA\,J1018 & { 23.37$\pm$0.02} & { 26.4$\pm$0.1 }&{ 27.2$\pm$0.1} & 1.8$\pm$0.1 & 2.4$\pm$0.2 & {  3.2}$\pm0.4$ & 1.5$\pm$0.1 & 0.8$\pm$0.1 & 1.0$\pm$0.1\\

XLSSC 122 & { 22.77$\pm$0.01} & { 26.1$\pm$0.1 } &{ 27.5$\pm$0.1 } & 2.2$\pm$0.1 & 2.6$\pm$0.1 & 2.8$\pm$0.4 & 1.2$\pm$0.1 & 1.4$\pm$0.1 & 1.6$\pm$0.2\\
 
\hline
\end{tabular}
\end{table*}

\subsection{Quantifying the intracluster light in z\texorpdfstring{$\sim$}{}2 proto-clusters}
To isolate the diffuse light in these proto-clusters we mask all the high surface brightness sources except for the BCG. We use the {\sc sextractor}-derived source catalogue described in section\,\ref{sec:skysub} and mask all objects, except the BCG, to four times the semimajor and semi-minor axes of objects using the {\sc sextractor} parameters ({\sc a\_image} and {\sc b\_image}). We checked the resulting masked images by eye and increased the mask size, by up to 10 times the semi-major and semi-minor axis, for large galaxies and very bright stars for which the smaller mask was insufficient. The masked galaxies in the 100 kpc surrounding the BCGs can be seen in Figure \ref{ICL}. In this figure we also see diffuse low surface brightness emission extending up to 100\,kpc from the BCG which resembles intracluster light.

We quantify the amount of light in the proto-cluster cores by measuring the luminosity of the BCG and intracluster light within three projected annuli of <10 kpc, 10 to 50 kpc and 50 to 100 kpc around each BCG (marked on Figure \ref{ICL}) and list them in Table \ref{tab:lum}. The inner-most aperture is likely to be dominated by light from the BCG whilst the outer annulus is likely to be dominated by intracluster light \citep{Joo_Lee_2023}. The uncertainty in the background within each of these apertures was estimated by placing apertures, of the same size and shape, at random locations over the fully masked image described in the section above on sky subtraction and image depth. The standard deviation of the fluxes (within the $\sim$1000 apertures which have a comparable unmasked area as the regions of interest) is taken as the  uncertainty in the sky background. 

We find that both proto-clusters contain at least two times more light in the region beyond 10 kpc than within 10 kpc of their central galaxies, demonstrating that these proto-clusters host significant amounts of intracluster light. We also measure the total light within a 100 kpc-radius circular aperture, and defined the concentration of 
 light as the luminosity within 10 kpc compared to the luminosity within a 100 kpc radius, with uncertainties resulting from the fractional errors in the 10 and 100kpc apertures added in quadrature. We find the light concentration in the centre of the XLSSC 122 proto-cluster is 
0.29$\pm$0.03, and 0.23$\pm$0.02 in the CARLA\,J1018 proto-cluster.

\begin{figure*}
\includegraphics[width=2\columnwidth]{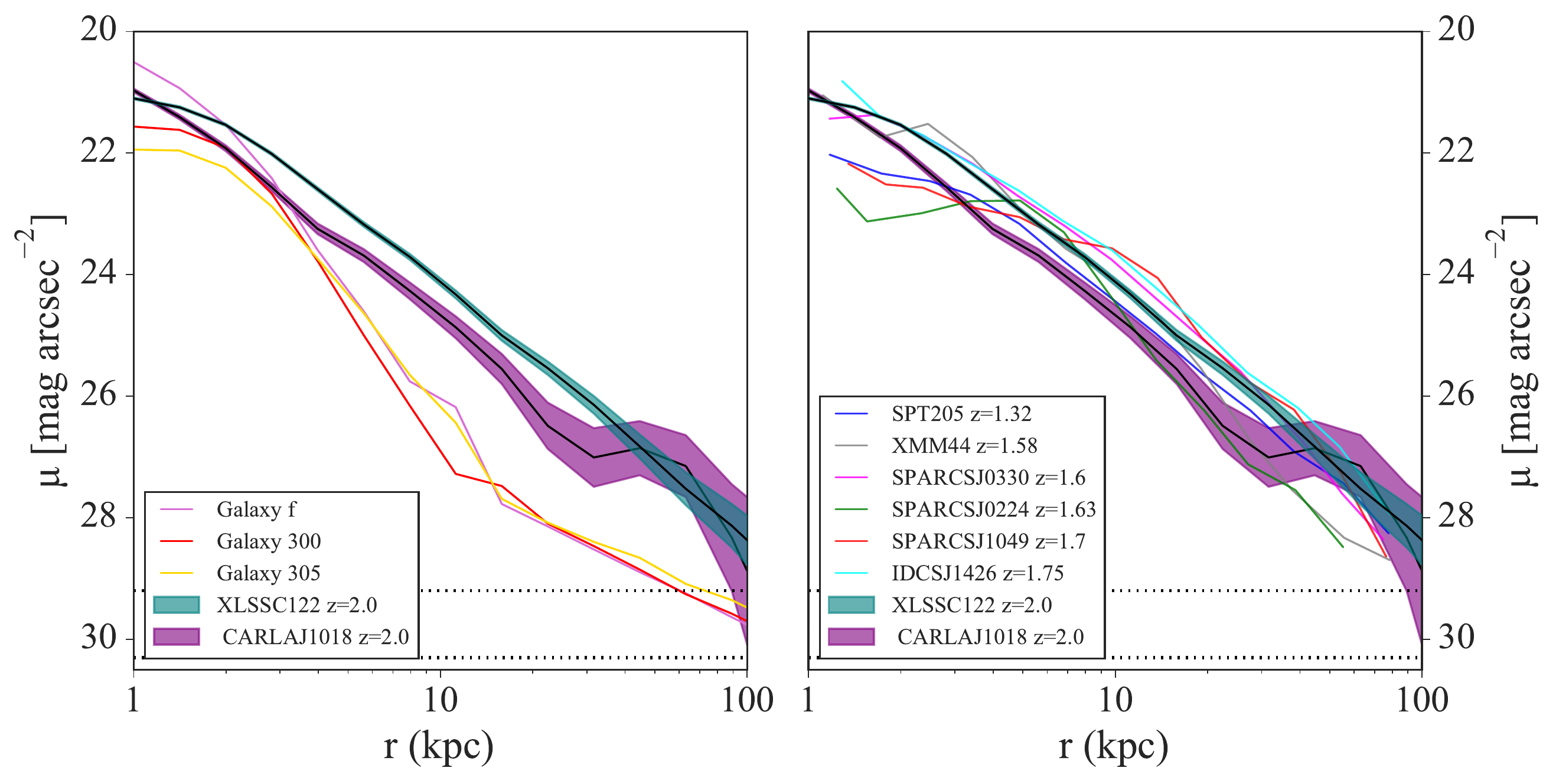}
\caption[parameters]{Radial surface brightness profiles for the BCG within the CARLA\,J1018 (purple) and XLSSC 122 (blue) proto-clusters. All other galaxies were masked from the $F140W$ {\it HST} images. {  The horizontal dotted lines at the bottom of the figures show the $3\sigma$ image depths for  CARLA\,J1018 (top) and XLSSC 122(bottom).} In the left panel, we compare the surface brightness profiles of the BCGs to the three most massive galaxies within the proto-clusters that lie far outside the cores ({\it \# 300} and {\it \# 305} from XLSSC 122, \citealt{Noordeh_2021}, and galaxy {\it f} within CARLA\,J1018). The BCGs have a uniquely extended morphology that is not shared by other proto-cluster galaxies. In the right panel, the surface brightness profiles are compared to BCGs at lower redshifts \citep{Demaio_2020}; redshift-corrected to $z=2$  to enable a direct comparison with the two proto-clusters at z$\sim$2. }
\label{sb}
\end{figure*}

We check whether the presence of diffuse light is unique to the BCGs or is a generic feature of massive proto-cluster galaxies by comparing the projected circular radial profiles of the BCGs to other proto-cluster galaxies with similar stellar masses of more than $\mathrm{10^{11}}$\,\Msun. We only find three galaxies with such high masses outside the central 100 kpc of the proto-clusters (galaxies labelled {\it \#300} and {\it \#305} in XLSSC 122 by \citealt{Noordeh_2021} and galaxy {\it f} from CARLA\,J1018). The surface brightness profiles of these three galaxies, shown in the left panel of Figure \ref{sb}, are much steeper than the BCGs, and the light concentration of these galaxies are all greater than 0.89 which is three times higher than either of the BCGs. The BCGs, therefore, have a distinct light profile that is more extended than other massive proto-cluster galaxies. 

In the right panel of Figure \ref{sb}, we compare the light profile of the $z\sim2$ BCGs to BCGs at $z\sim1.5$ that are known to be surrounded by intracluster light \citep{DeMaio_2019}. {  \citet{Joo_Lee_2023} performed multi-S\'ersic fitting to many of these profiles and found that light in the inner 10\,kpc was dominated by the BCG, whereas light beyond 10\,kpc was either dominated by the intracluster light component, or dominated by an extended, flat bulge component of the BCG up to at most 30\,kpc, and then intracluster light dominated beyond. The surface brightness profiles of the proto-cluster BCGs have remarkably similar profiles to the $z\sim1.5$ cluster BCGs beyond 10\,kpc, which we interpret as evidence that intracluster light is already present in these proto-clusters. }

\begin{figure*}
\includegraphics[width=1\columnwidth]{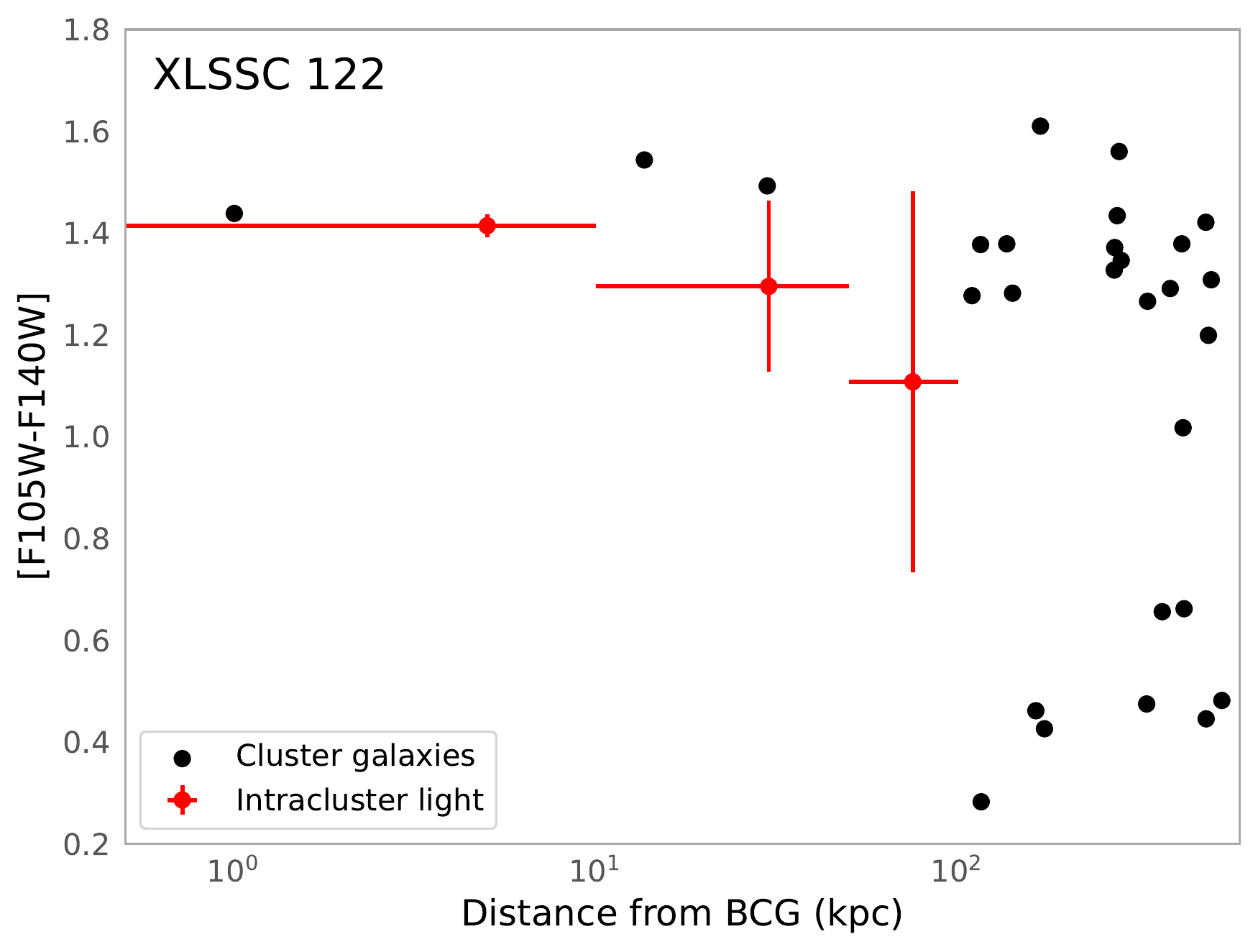}
\includegraphics[width=1\columnwidth]{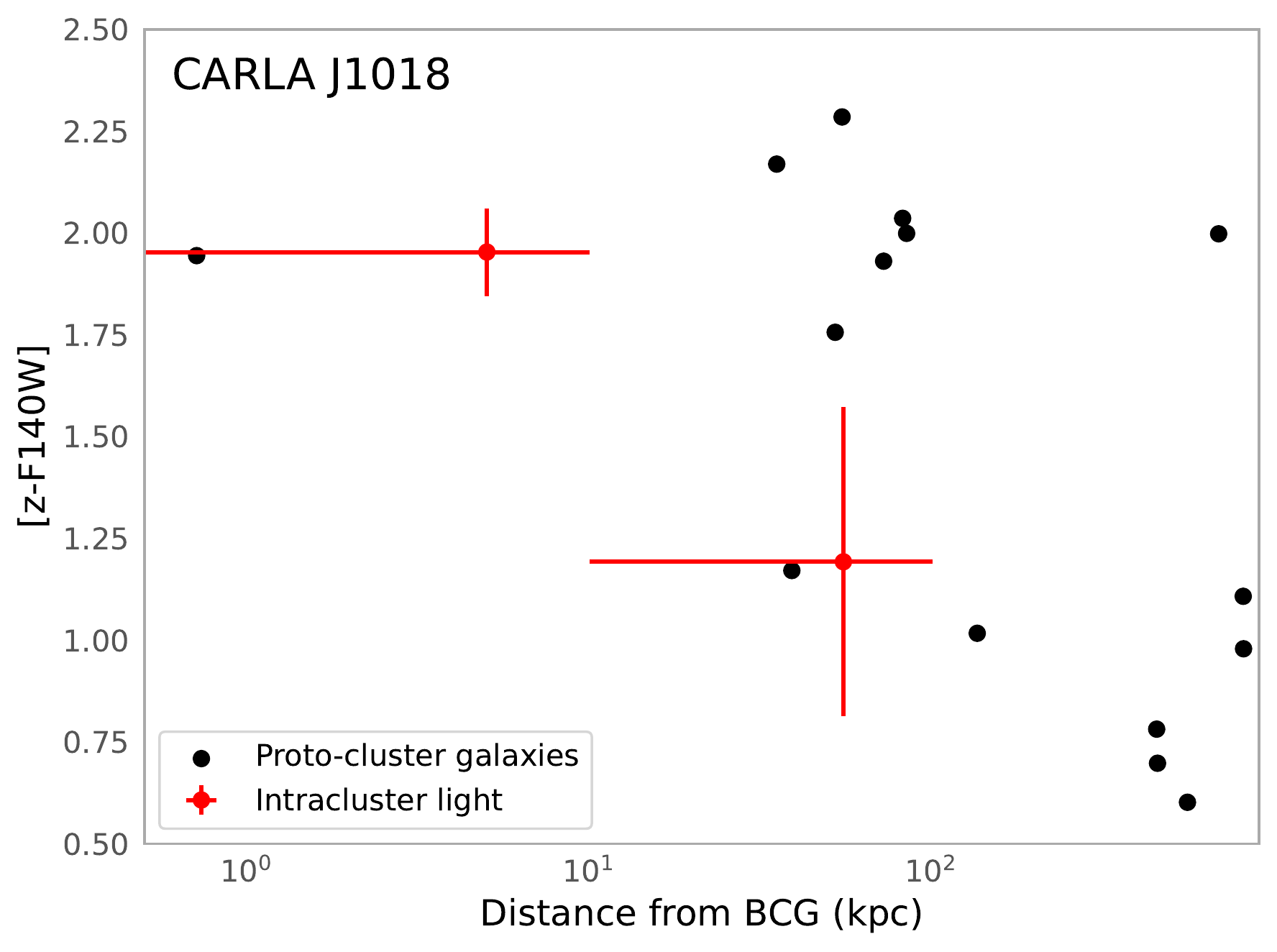}
\caption[parameters]{The colour of the BCG and intracluster light (red) compared to the proto-cluster galaxies (black dots) in XLSSC 122 (left) and CARLA\,J1018 (right). The intracluster light in XLSSC 122 has a similar colour to the red proto-cluster galaxies and we assume that the intracluster stars have the mass-weighted mass-to-light ratio of protocluster galaxies with colours of $[F105W-F140W] > 1.1$. The colour of the intracluster light in CARLA\,J1018 is significantly bluer in the outskirts than in the core region. We therefore calculate the mass of the stars within 10 kpc using the mass-weighted mass-to-light ratio of protocluster galaxies with colours of $[z-F140W] > 1.5$, whilst the mass of the stars beyond 10 kpc are calculated using the mass-weighted mass-to-light ratio of protocluster galaxies with colours of $[z-F140W] < 1.3$.}
\label{colors}
\end{figure*}

\begin{figure*}
\includegraphics[width=0.9\columnwidth]{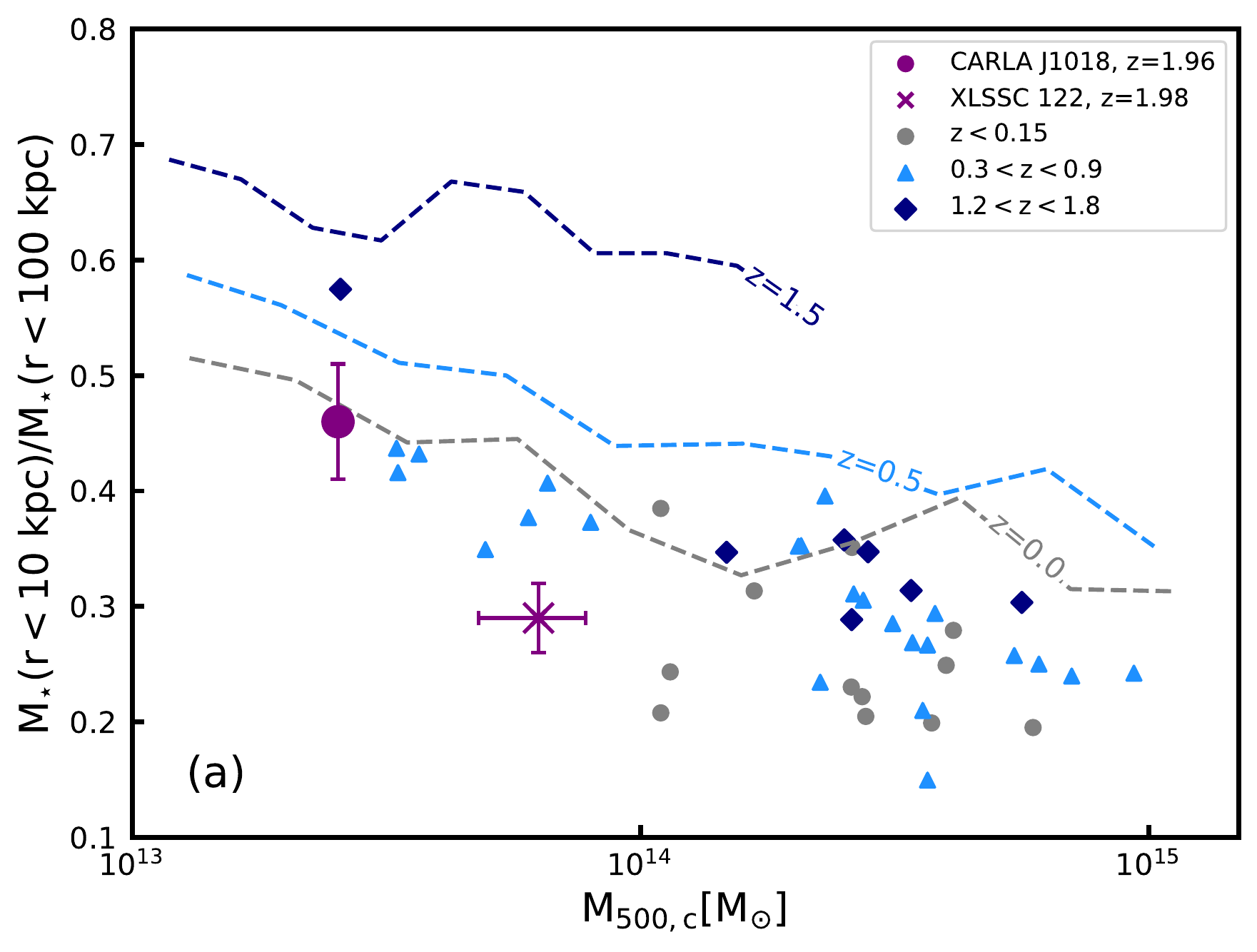}
\includegraphics[width=0.9\columnwidth]{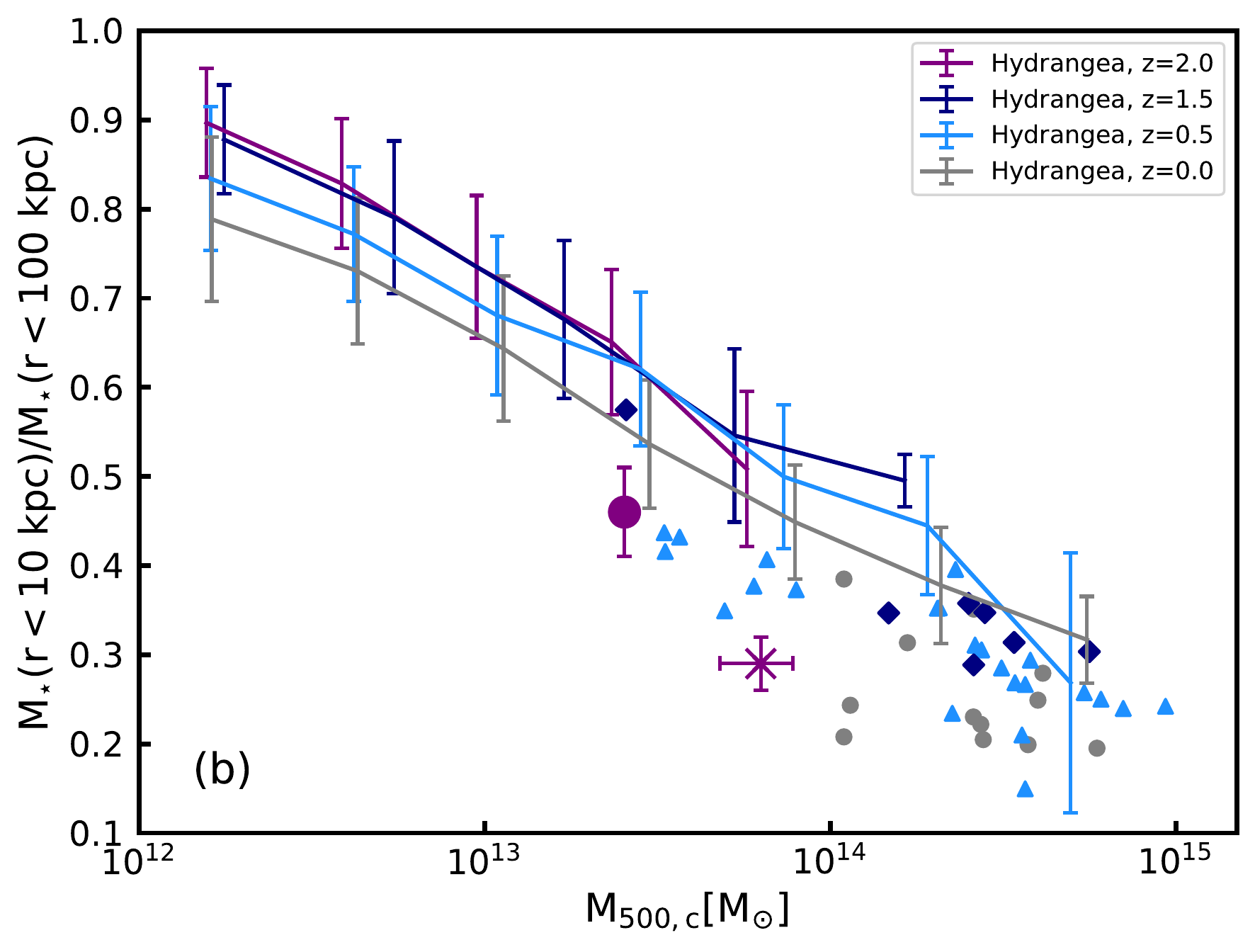}
\caption[parameters]{The concentration of stellar mass within the z$\sim$2 proto-clusters (purple symbols), defined as the ratio of stellar mass within 10 kpc to the stellar mass within 100 kpc, versus the total halo mass ($\mathrm{M_{500,c}}$. Grey and blue symbols show lower redshift clusters gathered from the literature \citep{Demaio_2020}, which suggests an inverse relationship between the stellar mass concentration of the core and the total cluster mass ($\mathrm{M_{500,c}}$). In the left panel, the dashed lines display the predicted concentrations from the semi-analytic models of \citet{Contini_2021}, whilst in the right panel the solid lines display the predicted concentrations from the Hydrangea hydrodynamical simulations \citep{Bah_2017}.}
\label{halo_mass}
\end{figure*}

\subsection{Stellar mass and concentration of the BCG and intracluster stars}

To compare our discovery of intracluster light within proto-clusters to the predictions of simulations, such as \cite{Contini_2021}, we need to convert the luminosity to an estimate of the stellar mass. We use the colour of the light to estimate the mass-to-light ratio in each annulus. We combine the {\it HST} $F140W$ images with images of light from below the Balmer break in the rest-frame of the proto-cluster galaxies. For XLSSC 122 we use an {\it HST} image taken through the $F105W$ filter, whilst for CARLAJ1018 we use a $z$-band image taken from the ground (from VLT). We estimate the colour of the light in three radial bins for XLSSC 122 and two wider bins for the shallower CARLA\,J1018 images and compare them to the colours of the proto-cluster galaxies in Figure \ref{colors}.

The light within 100\,kpc of the BCG of XLSSC 122 has a similar $[F105W]-[F140W]$ colour as the proto-cluster members that reside on the red sequence ($[F105W]-[F140W]>1.1$). We assume that the intracluster light is produced by the stripping and destruction of proto-cluster galaxies, and that all galaxies are stripped equally. Under these assumptions, it is appropriate to use the mass-weighted mass-to-light ratio of all the XLSSC-122 proto-cluster galaxies with $[F105W]-[F140W]>1.1$. Using the stellar masses (renormalised to a Chabrier IMF) listed by \cite{Noordeh_2021}, we find that such red galaxies contain a total mass of $\mathrm{10^{12}}$\,\Msun. We transform the observed fluxes into absolute magnitudes to derive a total $F140W$ luminosity of 1.8$\times\mathrm{10^{12} L_{\odot}}$. Thus, the observed mass-to-light ratio through the $F140W$ filter is 0.56 for the light in XLSSC 122.

The colour of the light within the core of CARLA\,J1018 varies with radius. To compare the {  colour of the diffuse} light  with the colour of the galaxies, we perform aperture photometry on the $F140W$ and $z-$band images using four times the semi-major and semi-minor axis of the catalogues detected by SExtractor for the masking. These large apertures are chosen to ensure we have all the galaxies’ light, even in the ground-based $z-$band image, which has a larger PSF than the space-based $F140W$ image.

In the inner 10\,kpc, the light matches the colour of the red sequence galaxies with $[z-F140W]>1.7$. The mass-weighted mass-to-light ratio for these galaxies is 0.89, which we adopt for the light within this aperture. This ratio is larger than that found for XLSSC 122 because XLSSC 122 contains several bright, low mass galaxies which bring down the mass-averaged mass-to-light ratio of the red galaxies. The CARLA\,J1018 $F140W$ and grism data is shallower than the XLSSC 122 data which results in fewer detections of low-mass red galaxies. However, any undetected low-mass galaxies are unlikely to greatly influence the mass-averaged mass-to-light ratios: the core region of CARLA\,J1018 contains several $\mathrm{10^{11}}$\Msun\ galaxies which negates the impact of any (undetected) low-mass galaxies on the mass-averaged mass-to-light ratios.

Beyond 10 kpc, the diffuse light of CARLA\,J1018 is significantly bluer and is similar to galaxies $i$ and $j$. The other five blue galaxies do not have measurable masses as they are not detected in the images taken by the VLT, and therefore they cannot be included in this calculation. The mass-weighted mass-to-light ratio of galaxies $i$ and $j$ is 0.32.

We convert the light measured in each aperture to stellar mass using these mass-to-light ratios and list them in Table\,\ref{tab:lum}. The main limitation of these results is the strong dependence of the stellar concentration on the assumed mass-to-light ratios for each radial bin. Simulations suggest that the intracluster light comes from stripping of the massive galaxies in the forming cluster \citep{Merritt_1984, Rudick_2006}, which justifies our using the mass-weighted mass-to-light ratios of the galaxies to estimate the mass-to-light ratio of the intracluster stars. However, to be cautious we explore how alternative mass-to-light ratios affect our conclusions in section\,\ref{sec:mass2light}. 

We define the concentration of stellar mass as the fraction of stellar mass within 100 kpc that lies within the central 10 kpc, $\mathrm{\frac{M_{*}(< 10 kpc)}{M_{*}(< 100 kpc)}}$, and find that the stellar mass is more centrally concentrated in CARLA\,J1018, with $\mathrm{\frac{M_{*}(< 10 kpc)}{M_{*}(< 100 kpc)}}= 0.46\pm0.05$, compared to XLSSC 122, with $\mathrm{\frac{M_{*}(< 10 kpc)}{M_{*}(< 100 kpc)}}= 0.29\pm0.03$. The quoted uncertainties are only from uncertainties in the light measurement within each aperture and do not include any systematic uncertainty from the assumed mass-to-light ratio. In Figure \ref{halo_mass}, we compare these stellar mass concentrations of the proto-cluster cores with lower redshift clusters from the sample of \citet{Demaio_2020}.  

The stellar mass concentration of the lower redshift clusters exhibits a mild trend with halo mass and no dependency on redshift. XLSSC 122 lies on the same relation of concentration-halo mass as the lower redshift clusters, and we use this relation to estimate a halo mass of CARLA\,J1018 based on its stellar mass concentration. To find a relation between $\mathrm{\frac{M_{*}(< 10 kpc)}{M_{*}(< 100 kpc)}}$ and $\mathrm{M_{500,c}}$, where $\mathrm{M_{500,c}}$ is the mass within the radius at which the density is 500 times the critical density of the Universe, we applied a least-squares algorithm to all clusters in the sample of \cite{Demaio_2020}, regardless of their redshift. We derive the following relation: $\mathrm{\frac{M_{*}(< 10 kpc)}{M_{*}(< 100 kpc)} = 2.57 - 0.16 \times log M_{500,c}}$. The stellar concentration of 0.46 within CARLA\,J1018 implies an approximate total cluster mass of $\mathrm{M_{500,c}\sim10^{13.4} M_{\odot}}$. We note this is within a factor of five of the halo mass estimate of \citet{Mei_2022} based on the galaxy overdensity, but both estimates should be considered unreliable as they have been derived through methods that have not been well tested.

\subsection{Comparing the proto-cluster stellar mass concentration to simulations}

The concentration of stellar mass in the central 100\,kpc of these proto-clusters provide us with a well-defined quantity that we compare to two state-of-the-art simulations of intracluster light 
formation. In Figure \ref{halo_mass}a we compare our data to the predictions of intracluster light from the semi-analytic model of \citet{Contini_2021}.  The intracluster stars in this model are produced from an analytic model that takes a fraction of the stars from satellite galaxies as they orbit the BCG, as well as harvesting a fraction of the stars from galaxies merging with the BCG. Full details of the model are provided in \cite{Contini_2014}. 

Once stripped, the intracluster stars are assumed to follow an NFW profile \citep{NFW_1997} adapted such that the concentration parameter of the intracluster light, $\rm C_{ICL}$, is modulated by a constant, ${\gamma (z)}$. By varying the stellar concentration parameter, these simulations can reproduce the observed stellar mass within 100 kpc of the centre of clusters at $0 < z < 1.5$ \citep{Contini_2020}. We use the optimal concentrations derived by \cite{Contini_2020}: ${\gamma (0) = 3}$, and  ${\gamma (>0) = 5}$, for the lines shown in Figure \ref{halo_mass}a. 

These models are designed to match the observed slow growth rate of the BCG, so in this simulation the BCG has mostly assembled by $z\sim 1$ and the surrounding intracluster light builds up around it over the following 7\,Gyr. The stellar concentrations therefore decrease with decreasing redshift as stars from mergers and stripping preferentially accumulate beyond the outskirts of the BCG. The highest redshift model predictions are at z$ \sim 1.5$, since above this redshift the intracluster light is predicted to be negligible and the stellar concentrations tend towards unity. 

\begin{figure}
\includegraphics[width=1\columnwidth]{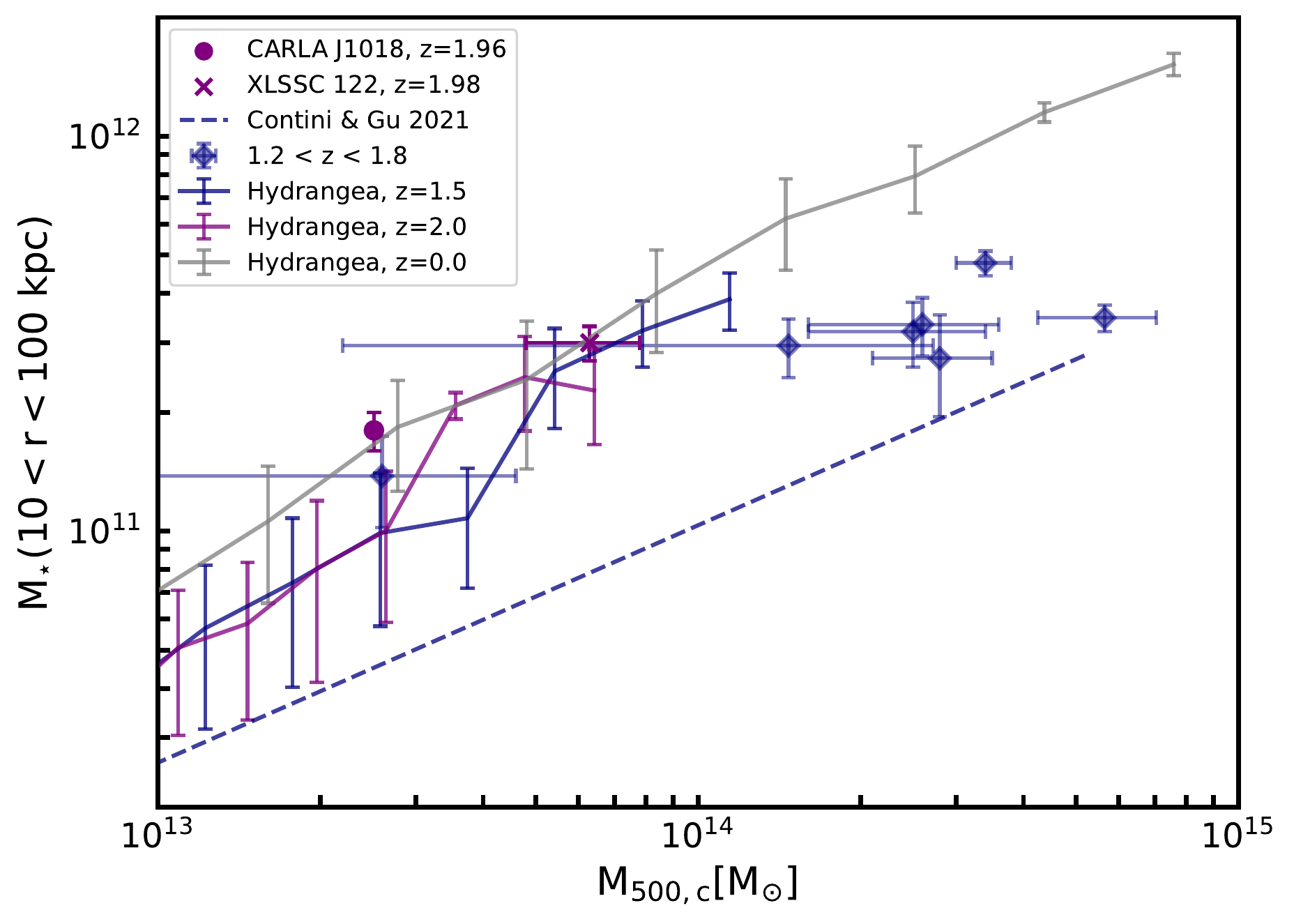}
\caption[p]{  The amount of stellar mass in the annulus $10 <R<100$\,kpc around the BCGs in the $z=2$ proto-clusters (purple symbols), and $z\sim1.5$ clusters from \citep[][blue diamonds]{Demaio_2020}, versus the total halo mass, $\mathrm{M_{500,c}}$. The dashed blue line displays the predicted concentrations from the semi-analytic models of \citet{Contini_2021} at $z=1.5$, and the solid lines display the predicted concentrations from the Hydrangea hydrodynamical simulations \citep{Bah_2017} at three different redshifts. The Hydrangea simulations are in reasonable agreement with the observations in the overlapping mass range ($\mathrm{M_{500,c}}<10^{14}$\Msun), whereas the semi-analytic models underpredict the amount of intracluster stars at these high redshifts.}
\label{amount}
\end{figure}

The \citet{Contini_2021} predictions agree  reasonably well with the observed concentrations at $z = 0$ and $0.5$, although the predictions are higher than most of the observations. The main discrepancy occurs at $z=1.5$ where the models have significantly higher concentrations than both $z\sim 2$ proto-clusters as well as other clusters at $z\sim$1.5. This remains true regardless of the assumed halo mass of CARLA\,J1018. 
As shown fig.\,1 of \citet{Contini_2021}, and reproduced in Fig.\,\ref{amount}, these models underpredict the mass of intracluster stars at $z=1.5$ within the $10 <R<100$\,kpc annulus surrounding the BCGs, which explains why the predicted concentrations are much higher than  observed. Our discovery of BCGs at $z\sim 2$ with low stellar concentrations and significant stellar mass at $>10$\,kpc implies that the formation of the intracluster light began earlier than predicted by this model.  

We next compare the observed stellar mass concentrations to the Hydrangea hydrodynamical simulations \citep{Bah_2017}. These are a suite of 24 zoom-in simulations of massive galaxy clusters within which the equations of gravity are solved for collisionless dark matter and stellar particles, with additional hydrodynamical equations solved for gas particles. Numerical subgrid algorithms are used to solve for the other ingredients of galaxy formation, such as cooling, star formation, stellar feedback and black hole growth (see see \citealt{Schaye_2015} and \citealt{Bah_2017} for details). 

The positions of star particles are traced throughout the simulation so the stellar distribution can be used to test the fidelity of the model. Using these simulations, \cite{Asensio_2020} showed that the distribution of the intracluster stars closely follows that of the total mass distribution at $z\sim 0$. This agrees with recent observations that showed the intracluster light followed the mass distribution within the central 140\,kpc  \citep{Montes_2018}. However, we caution that these simulations produce too many stars in the centres of clusters and the BCGs end up three times too massive by the present day \citep{Bah_2017}. This means that the galaxy model has not yet solved the issue of restraining the BCG stellar growth.

To create the data shown in Figure \ref{halo_mass}b we extracted the stellar mass concentration within four snapshots of the simulations: $z = 0.0$, $z=0.47$, $z=1.49$ and $z=1.99$. Within each snapshot, we identified the central galaxy (the BCG) of each dark matter halo with a mass greater than $10^{12}$ \Msun. The stellar mass concentration is calculated as the ratio of stellar mass within {  a projected radius of} 10\,kpc to the stellar mass within {  a projected radius of} 100\,kpc, excluding any stars within these {  2D} apertures that are gravitationally bound to satellite galaxies. We separated the data from each snapshot into halo mass bins, then calculated the median stellar mass concentration and the standard deviation of the concentration for the halos in each halo mass bin, which are displayed in Figure \ref{halo_mass}b.

In these hydrodynamical simulations the redistribution of stars during mergers and stripping depends only on gravity and the position of stars within galaxies. Hence the distribution of BCG and intracluster stars in this simulation is a direct prediction of hierarchical merging, modulated only by the galaxy evolution model. 
 
We find that the concentrations from these simulations are in reasonable agreement with the observations  at $z=0$ and 0.5, although the simulations tend have higher concentrations than the observations, similar to the semi-analytic models mentioned above. 
 
In contrast to the semi-analytic models of \citet{Contini_2021}, the concentrations of the hydrodynamical simulations at $z>0.5$ have little further evolution with redshift in the $M>10^{13}$\Msun\ mass regime, and  only minor evolution at lower masses. In general, the  hydrodynamical simulations predict that the stellar concentration depends on the halo’s total mass, but has relatively little dependency on redshift, especially at $z>0.5$.  Thus, there is no redshift at which the model does significantly worse at reproducing the observations; the stellar light is slightly more concentrated in the simulated haloes than in the observations at all redshifts probed.
 
\citet{Bah_2017} has, however, shown that the simulations produce BCGs that are too massive which can impact the stellar concentrations. Therefore we also compare the predicted and observed stellar mass in the annulus $10 <R<100$\,kpc surrounding the BCGs in Fig.\,\ref{amount}. The Hydrangea simulations predict a similar amount of intracluster stars to the observations at $\mathrm{M_{500,c}}<10^{14}$\Msun, but  unfortunately do not probe haloes of masses $\mathrm{M_{500,c}}>10^{14}$\Msun\, where most of the $z\sim1.5$ observations exist. The stellar mass within this region of the two proto-clusters lies on the upper bound of the predictions, but within the scatter. 
 
The Hydrangea simulations, therefore, predict a level of intracluster light that is consistent with the observations at $z=2$, but the concentrations are higher than observed because the simulations produce too much mass in the central 10\,kpc of the BCG.  We therefore still lack a simulation model that can predict realistic amounts of intracluster light at $z=2$ while also restraining excessive BCG growth.

\vspace{-0.5cm}

\section{Discussion}
\begin{figure}
\includegraphics[width=1\columnwidth]{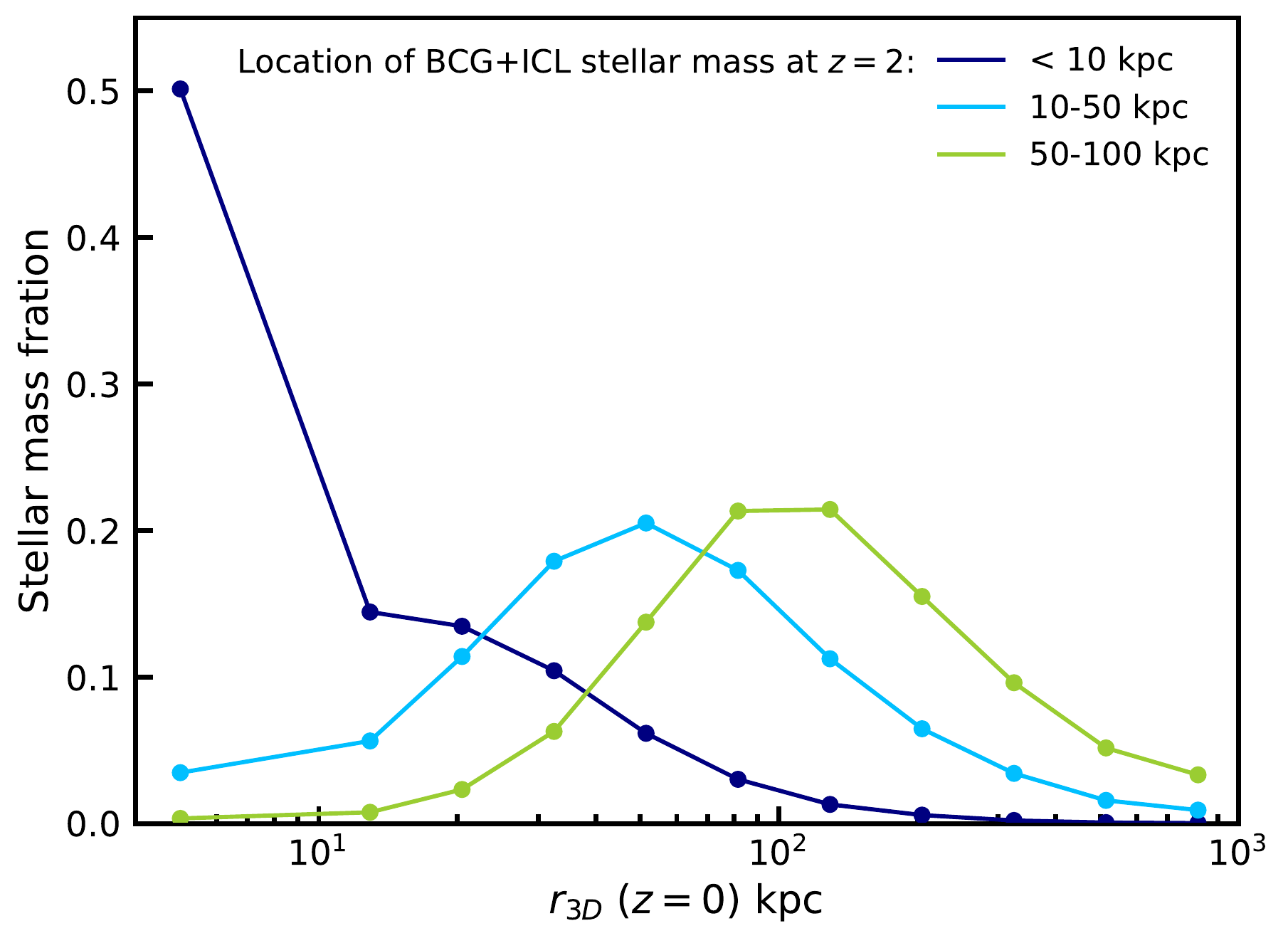}
\caption[ending]{{ The location where the stellar mass of the Hydrangea proto-clusters' BCG and intracluster light ends up by $z=0$. The coloured lines show the $z=0$ location of stars that were located at $z=2$: within 10\,kpc of the BCG centre (dark blue); between 10 and 50\,kpc (light blue); and between 50 and 100\,kpc (green).}}
\label{where_end}
\vspace*{-0.3cm}
\end{figure}

\subsection{The implications of finding intracluster light in proto-clusters}
Our work extends the baseline over which intracluster light has been investigated to $z\sim2$. At this epoch, typically less than 20\% of the matter that will end up in the cluster has assembled into the main halo \citep{Muldrew_2015}, and the halos that we have observed are below the general limit of a cluster-sized halo and more akin to galaxy groups. But our images show that the central 100\,kpc of these proto-clusters already contain intracluster stars.

The Hydrangea simulations agree that intracluster light can exist at such a high redshift, and go further to suggest that intra-halo stars, i.e. stars that are in halos but gravitationally unbound from any galaxy, are present in massive haloes up to at least $z\sim2$. As these massive haloes merge to form larger groups and accrete onto clusters their intra-halo stars will combine. Stars that were already unbound from galaxies when they accrete onto the cluster are defined as having been \lq pre-processed\rq. Since the Hydrangea simulations suggest that all massive halos contain intra-halo stars, it is likely that pre-processed stars are a significant source of intracluster stars. Therefore semi-analytic simulations should take care to include intra-halo stars at $z>1$, even in halos of masses as low as $10^{12}$\Msun. 

Our findings also argue against a significant contribution of intracluster stars within the central 100\,kpc coming from the stripping of satellite galaxies as they orbit in the cluster. Such a stripping mechanism would build up intracluster light over time, regardless of whether the halo grows significantly in mass. Thus, if stripping of orbiting satellites were a dominant contributor to intracluster stars we would expect the stellar mass concentration to strongly depend on redshift, and only weakly depend on halo mass. Such evolution in the stellar mass concentration can be seen in the models of \citet{Contini_2021} displayed in Figure \ref{halo_mass}a, but these do not match the trends seen when combining our data with the larger sample from \citet{Demaio_2020}. 

Our observational data find the opposite trends: a dependence on halo mass and little (or no) dependence on redshift. The stellar mass concentration in the $z\sim2$ proto-clusters is comparable to group-sized halos at $0.1<z<0.9$ which have similar masses. The groups at $z<0.9$ have had much more time to strip stars from satellites than the $z\sim2$ proto-clusters, thus such stars cannot be a major source of intracluster stars within the central 100\,kpc. It is plausible, however, that such stars contribute much more to the intracluster light at larger radii.

These results support the recent finding of \citet{Ko_Jee_2018} and \citet{Joo_Lee_2023} which revealed massive clusters at $z\sim1$ that contained $\sim17$\% of the total stellar light of the clusters within the diffuse intracluster component, which is similar to the fraction of intracluster light in local clusters. These authors suggested that the dominant form of ICL production could be through the accretion of pre-processed stray stars, and our observations provide evidence of an abundance of such stars in proto-clusters.

\subsection{The fate of the BCG and intracluster stars found in proto-clusters}

The amount of intracluster stars in the proto-clusters is likely to increase over time because the proto-clusters are likely to evolve into clusters that are an order of magnitude more massive \citep{Muldrew_2015}. As we can see from Fig.\,\ref{amount}, such massive clusters are predicted to have significantly more stellar mass in the $10 <R<100$\,kpc region than their low mass progenitors. Furthermore, local clusters are known to have significant intracluster light at more than 100\,kpc distant from the BCG. 

In Figure \ref{where_end} we explore the fate of the stars within 100\,kpc of the $z=2$ BCGs that were identified in the $M>10^{13}$\Msun\ Hydrangea haloes. We first divide the stellar population at $z=2$ into three radial bins: $<10$\,kpc, between 10 and 50\,kpc, and between 50 and 100\,kpc. We then plot the fraction of stellar mass within each of these bins that ends up at various radii within the cluster at $z=0$. Since we want to know where the $z = 2$ stars end up, the stellar mass of each stellar particle is always taken as the one measured at $z = 2$. By $z = 0$, however, stellar mass loss will reduce the mass of the stellar particles, but this does not change the profiles significantly.

Over time the $z=2$ stars diffuse from their initial position with most stars moving to larger radii. Only half of the stars that lay within the central 10\,kpc of the BCG remain within the same region. The other half move further out and a few percent reach distances of more than 100\,kpc. Similarly, approximately half of the stars in the $10<R<50$\,kpc bin remain in the same location and the other half move to larger distances; almost a quarter of the stars travel to distances larger than 100\,kpc. The largest amount of movement occurs in the $50<R<100$\,kpc bin, where only $\sim30$\% of the stars remain in the same bin by $z=0$ and over 55\% move further out beyond 100\,kpc and several percent reaching distances of over 400\,kpc. Overall, by $z=0$ the stars end up further away from the BCG than at $z=2$.

Spread over such a large area, the resulting surface brightness from the stars at $R>100$\,kpc would likely be below current measurements of nearby clusters.  Therefore, further production of intracluster stars or accumulation of previously stripped stars from infalling groups 
(e.g.~\citealt{Bahe2019}) must occur to result in the prominent intracluster light seen at several 100\,kpc from the BCGs in present day massive clusters. Such growth in the intracluster light is expected given the results of \citet[e.g.][]{Burke_2012} and \cite{Demaio_2020} who showed that the stellar mass in clusters grows from the inside-out. \citet{Burke_2012} showed that the low-surface brightness light ($\mu_J > 22$\,mag\,arcsec$^{-2}$)  in the ${R<R}_{500}$ region of $z=1$ clusters grows by $2-4$ times by the present day, whereas \cite{Demaio_2020} showed that the amount of stellar mass within the central 100\,kpc of clusters continued growing until $z\sim0.4$. These works suggest the intracluster light beyond 100\,kpc will likely continue growing to the present day, and the mass growth rate of intracluster stars exceeds that of the halo. Therefore, the diffuse light we see in the proto-clusters is likely to grow both in size and mass over the following 10\,Gyr.

\subsection{Prospects for detecting proto-cluster intracluster light with ESA’s {\it \textbf{Euclid}} Mission}

Our discovery of intracluster light at $z\sim2$ means that it is possible to use the signature of diffuse low surface brightness light and low stellar mass concentrations to identify clusters and proto-clusters from high-resolution images, such as will be provided by ESA’s {\it Euclid} mission. This is particularly useful at $z>1.5$, where the photometric redshifts that {\it Euclid}'s cluster finders use are not precise enough to identify galaxy clusters. The presence, or lack, of intracluster light can be used to remove contaminants from the cluster catalogues based on photometric redshifts. Furthermore, since the presence of diffuse light is unique to the central galaxies in proto-clusters, this is an excellent way to identify the dominant halo within a proto-cluster.

The {\it Euclid} Wide Survey is predicted to reach a surface brightness of 29.1 mag/$\mathrm{arcsec^2}$ in VIS (3$\sigma$ over 100 $\mathrm{arcsec^2}$) and 27.7 mag/$\mathrm{arcsec^2}$ for Y, J, and H$-$band NISP images \citep{euclid_2022, euclid2_2022}, whilst the Deep Survey will reach 2 magnitudes fainter. This means that the Deep Survey will reach similar depths to the {\it HST} $F140W$ images we use in this work ($29.2$ and $30.3$\,mag/$\mathrm{arcsec^2}$ in CARLA\,J1018 and XLSSC-122, respectively) but over a much larger area. 

It is also likely that intracluster light will be detected in the Wide Survey of {\it Euclid}. The surface brightness of the intracluster light in the region between 10 and 50\,kpc ($\sim$100 $\mathrm{arcsec^2}$) in both proto-clusters presented in this work is $[F140W]\sim 26.5$\,mag/$\mathrm{arcsec^2}$, and $[F140W]\sim 27.1$\,mag/$\mathrm{arcsec^2}$ averaged over the larger region between 10 and 100\,kpc ($\sim430\,\mathrm{arcsec^2}$). Thus, we predict that the intracluster light in both proto-clusters should be detected in the near-infrared images of {\it Euclid}'s Wide Survey.

Whether the intracluster light is also detected in the VIS images depends on the colour of the light. The reddest colour of the intracluster light that we measure is $[z]-[F140W]\sim2$, which means the surface brightness could be as low as $z=29.1$\,mag/$\mathrm{arcsec^2}$. The VIS instrument covers the wavelength range $550$ to $900$\,nm, so it covers light blueward of the $z-$band image we use in this work. Hence, intracluster light at $z \sim 2$ is likely to be detected in VIS images of the Deep Survey, but we are unlikely to detect intracluster light in the Wide Survey unless the intracluster stars are young and blue. We conclude that the NISP images are better suited to search for intracluster light in distance clusters and proto-clusters. 

Additionally, the intracluster light maps in Figure \ref{ICL} show that the light is not uniform and brighter regions of diffuse emission will appear well above {\it Euclid}’s detection limits. This will allow detailed measurements of the morphology and colours of the intracluster light.

{  The Vera C. Rubin Observatory (LSST) will provide complementary data in the optical bands to even greater depth than the {\it Euclid} survey. However, the prospects of detecting intracluster light in proto-clusters with LSST are limited for two reasons. First, the survey is optimised for depth in the optical bands and the $z$ and $y$ band images will be the shallowest. Therefore, the LSST survey will probe the rest-frame ultraviolet light from proto-cluster intracluster stars which will only be prominent when the intracluster stars are young. Second, the relatively poor spatial resolution of LSST data compared to {\it Euclid} means that it will be harder to separate galactic light from diffuse intracluster light in the dense core of proto-clusters.}

\subsection{The impact of the mass-to-light ratio on the stellar mass concentration}
\label{sec:mass2light}
The uncertainties on the colour, and hence mass-to-light ratio, of the intracluster light beyond 10 kpc within CARLA\,J1018, and beyond 50 kpc in XLSSC 122 can significantly affect our conclusions of a low stellar mass concentration, so it is worth investigating how different mass-to-light ratios will affect our conclusions. First, we consider an extreme case in which the mass-to-light ratio beyond 50 kpc of XLSSC 122 is extremely low such that the mass in this region is negligible. Even in this extreme case, the mass concentration would increase to only 0.46$\pm$0.04, which is still compatible with the trend for other clusters seen in Figure \ref{halo_mass} and lies far below the model predictions of \citet{Contini_2021}.

Next, we consider the lowest plausible value of the mass-to-light ratio for the ICL beyond 10\,kpc in CARLA\,J1018. We start by examining colour variations in the ICL. We place four rectangular apertures across the brightest region of the ICL in the $F140W$ image, but not covering the central 10\,kpc from the BCG. The centres of the apertures were at [$154.6255$, 5.5183], [154.6264, 5.5175], [154.6279, 5.5161], [154.6288, 5.5152], each of $39 \times 93$\,kpc and at an $45^{\circ}$ angle. All apertures had measurable $F140W$ and $z-$band fluxes at more than a 1.5$\sigma$ level. The colours of the four regions were $[z]-[F140W] = 0.9$, $1.1$, $1.0$ and $1.6$ from South-east to North-west, which means there is colour variation in the intracluster light and a single mass-to-light ratio does not capture the complexity of this system. Nonetheless, we can use a single mass-to-light ratio to estimate a minimum mass in this outer region and take that as a lower limit to the true mass. The colour of the intracluster light in the annulus of 10 to 100\,kpc is $[z]-[F140W] =1.2\pm0.4$. Thus, the minimum colour of the intracluster light, assuming maximal measurement uncertainties is  $[z]-[F140W] = 0.8$. The intracluster light in all smaller regions have a redder colour that this, reassuring us that this is the bluest possible colour of the intracluster light.

\begin{figure}
\includegraphics[width=1\columnwidth]{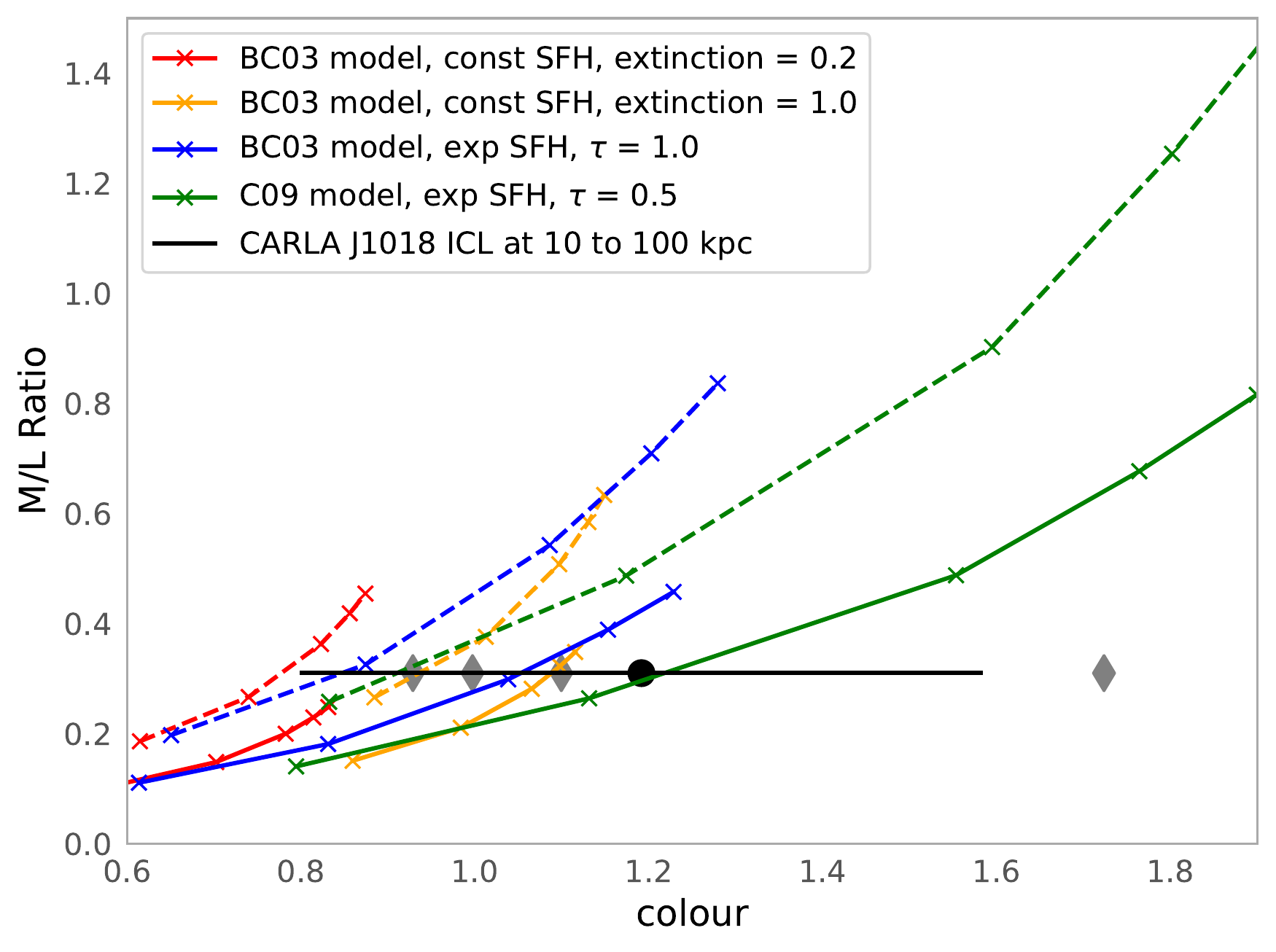}
\caption[parameters]{The relationship between the [$z - F140$] colour and mass-to-light ratio for galaxies at $z=1.96$. The colour of the intracluster light in CARLA\,J1018 between 10 and 100\,kpc is shown as the black point with $1\sigma$ uncertainties. The grey diamonds display the variety of intracluster light colour in the four rectangular regions described in the text. The coloured lines show the mass-to-light ratio from galaxies simulated with different star formation histories and Chabrier (solid) or Salpeter (dashed) initial mass functions. The crosses from the bottom left to top right of each line mark the colour of galaxies formed at $z=2.5$, 3, 4, 5 and 6 and observed at $z = 1.96$. The lowest possible colour of the intracluster corresponds to a minimum mass-to-light ratio of 0.14. We take this to be the minimum possible mass-to-light ratio for the intracluster light in CARLA\,J1018.}
\label{ML_ratio}
\end{figure}

We compare this colour to a set of stellar population models in Figure \ref{ML_ratio}. Such a blue colour can only be achieved with a stellar population model that is actively forming new stars, such as a constant star formation history or a very young (0.5\,Gyr), exponentially declining model with a long timescale for the decay of the star formation rate. The mass-to-light ratio also depends on the assumed IMF, with a Salpeter IMF producing a higher mass-to-light ratio than the Chabrier IMF. Throughout this work we assume a Chabrier IMF, however, evidence points to a peculiar stellar IMF in clusters \citep{Friedmann_2018}. By assuming a Chabrier IMF, we therefore take the minimum plausible mass-to-light ratio for the intracluster light. For a Chabrier IMF, a colour of $[z]-[F140W] = 0.8$ corresponds to a minimum mass-to-light ratio of 0.14. Applying this minimum mass-to-light ratio to the intracluster light beyond 10\,kpc, the luminosity concentration of 0.23 implies a mass concentration of 0.66. This value is still compatible with the observed trend for lower-redshift cluster BCGs, and below the measurements for other massive galaxies in the proto-cluster.

\subsection{Defining the BCGs in the proto-clusters}

The selection of the BCG can also affect our conclusions of a low stellar mass concentration. {  There is no other plausible candidate for the BCG in XLSSC~122 since the galaxy we have labelled as the BCG is 1\,mag brighter, and 7.6 times more massive than the second ranked galaxy in this proto-cluster \citep{Noordeh_2021}. On the other hand,} the selection of galaxy {\it a} as the BCG of CARLA\,J1018 is not certain as galaxies {\it b}, {\it c}, {\it d}, as well as the quasar, could all be the most massive galaxy in the proto-cluster. {  Galaxy {\it b} is also brighter than galaxy {\it a} in F140W, although both galaxies have consistent fluxes within uncertainties in the $K-$band, and as can be seen from Fig.\,\ref{spectra}, the SED fits predict that redward of 2-microns, galaxy {\it a} is likely to be brighter than galaxy {\it b}}. We therefore used the local surface density of galaxies to determine which of these galaxies is most likely to be at the barycentre of the proto-cluster. We selected galaxy {\it a} because it is the most massive and has the highest galaxy density on the scale of 100\,kpc out of all galaxies in the proto-cluster. 

{  Whilst these reasons justify our choice of galaxy {\it a} as the BCG, we also find that} our results do not qualitatively change if we chose galaxies {\it b}, {\it c} or {\it d} to be the BCG: we still find significant  diffuse light {  beyond 10\,kpc of the BCG core} regardless of which galaxy is chosen to be the BCG. Quantitatively, the amount of light within  the 100\,kpc radius would be reduced if the BCG was chosen to lie at the edge of the high-density region shown in the insert of Figure \ref{RA_DEC}, and the amount of {  diffuse} light that would be classified as residing beyond 100 kpc would increase. {  For example, if we selected galaxy {\it b} as the BCG, then we obtain a BCG+ICL observed F140W magnitude of $21.70\pm0.02$\,mag  within 10\,kpc and $20.9\pm0.2$\,mag in the 10 to 100\,kpc annulus. This results in a light concentration of $0.32\pm0.03$. This is slightly more concentrated than if we selected galaxy {\it a} as the BCG, but it is still far lower than the light concentration found for the other massive proto-cluster galaxies located outside the core.} Therefore the exact values of the mass and light concentrations will increase, although they are still low compared to other massive galaxies, both within and beyond the proto-cluster. Thus, our conclusion remains valid as long as the dense region in the insert of Figure \ref{RA_DEC} is the barycentre of the proto-cluster.

The argument given above only holds if we can use the surface density of galaxies as a proxy for the true 3D density; if the dense core region in the insert of Figure \ref{RA_DEC} is only due to the projection of unrelated galaxies, then this region may not be the barycentre of the proto-cluster. We therefore compare the probability that the galaxy overdensity is due to a chance line-of-sight  distribution to the probability that these galaxies are in close proximity and are interacting with each other. 

The area of the region depicted in the insert of Figure \ref{RA_DEC} is $0.31 \times 0.32$ co-moving $\mathrm{Mpc^{2}}$ ( $16.8 \times 17.1 \mathrm{arcsec^{2}}$) and the line-of-sight co-moving radial distance spanned by the interval $\Delta z = 0.04$ is 60.4\,Mpc, so the volume is $6.1 \ \mathrm{Mpc^{3}}$. The number density of $>\mathrm{10^{11}}$\,\Msun\ galaxies at $z\sim2$ is $10^{-4}$\,Mpc$^{-3}$ \citep{Muzzin_2013} so the probability of observing any galaxy more massive than  $10^{11}$\,\Msun\ in this volume is 0.06\%. The probability of detecting five unrelated galaxies, of more than $\mathrm{10^{11}}$\,\Msun, within this volume is vanishingly small at less than one in a trillion.

On the other hand, the chance of observing two massive galaxies that are in the process of merging is much higher. Simulations predict that a $\mathrm{10^{11}}$\Msun\ galaxy at $z = 2$ will merge with another galaxy, of at least a quarter of its mass, at a rate of 0.15\,Gyr$^{-1}$ \citep{Rodriguez_2015}. Galaxy mergers can be identified as close pairs of galaxies, and the length of time a merger is visible depends on the projected separation and mass ratio of the pair. For galaxies within a $1:4$ mass ratio this timescale varies from a few hundred Myr to 1.5\,Gyr depending on the projected separation \citep{Lotz_2011}. The projected separations of the galaxies in the core region range between 12 and 36\,kpc (proper) so these mergers will be visible for an average of 0.6 -- 1\,Gyr. Thus, the probability of observing a massive galaxy undergoing a merger with another nearby galaxy is 10 to 15\%. Since there are five galaxies more massive than $\mathrm{10^{11}}$\Msun\ in the core, and hence 10 unique pairs, there is a 60 to 80\% chance that we would observe these galaxies undergoing a merger. Assuming each merger is mutually exclusive and treating the closer merger pair as a single entity, the probability of observing five massive galaxies merging at the same time is 0.01 -- 0.05\%. This is much higher than the chance of observing 5 unrelated massive galaxies distributed across the 60.4 cMpc line-of-sight, hence it is more likely that the dense galaxy group we have observed in CARLA\,J1018 is the barycentre of the proto-cluster than a chance line of sight alignment. 

The space density of $>10^{11}$\Msun\  galaxies is $10^{-4}$\,Mpc$^{-3}$ \citep{Muzzin_2013}, and we have calculated that between one in a hundred to one in twenty of these will be undergoing extreme merging events such as that seen in the core of the CARLA\,J1018 proto-cluster. This means the space density of such extreme merging events is $10^{-8}$\,Mpc$^{-3}$. This is the same space density as galaxy clusters with masses greater than ${10^{15}}$\Msun\ in the nearby Universe \citep{Vikhlinin_2009}. It is therefore plausible that such extreme merging events signpost the formation of the core within the most massive  galaxy clusters.

\section{Conclusions}

We report on the detection of intracluster light within two proto-clusters at $z\sim2$ using deep {\it HST} near-infrared images. This extends our understanding of intracluster light to the numerous group-sized halos that exist at this redshift; previous measurements focused on the most extreme (and rare) cluster-sized halos at $z\sim1.5$ which may not be typical cluster progenitors. 

We identified the BCGs of the proto-clusters as the most massive galaxies within the densest regions of the proto-clusters, and measure the amount of diffuse light surrounding these galaxies. We found that the flux of diffuse light between 10--100\,kpc is more than double the flux from within 10\,kpc. We showed that this extended morphology is similar to the profiles of BCGs within massive clusters at $1.24<z<1.75$ which are known to host intracluster light. Furthermore, this profile differs from the other massive  galaxies in the proto-cluster, whose light profiles are at least a factor of 3 more concentrated than the BCGs. Based on these observations, we conclude that the proto-clusters contain significant intracluster light.

We used the colour of the intracluster light to estimate its mass-to-light ratio and calculate the concentration of stellar mass in the core of the proto-clusters.  We found that only a quarter to a half of the stellar mass within 100\,kpc is located within the central 10\,kpc.  We combine our data with that of \citet{Demaio_2020} to show that this low concentration is comparable to that found in similar-sized halos at lower redshifts, and such low concentrations at $z\sim2$ are in disagreement with the semi-analytic models of intracluster light by \citet{Contini_2021}. Our discovery implies that the formation of the intracluster light began earlier than the $z\sim1 - 1.5$ period predicted by these models.

To investigate how this intracluster light could form so early, we compared the stellar mass concentrations in clusters and proto-clusters to that of central galaxies of massive halos in the Hydrangea hydrodynamical simulations \citep{Bah_2017}. We found that these simulations agree with the trend in the observations: the stellar mass concentration depends on the halo mass, but does not depend on redshift. Hydrangea also predicts that intra-halo stars are ubiquitous in massive halos at all redshifts, and even found in halos with masses as low as $10^{12}$\Msun. We interpret these findings as evidence that pre-processed free-floating stars from accreted halos is a major contributor to the intracluster light within 100\,kpc of clusters and protoclusters, whilst few of the stars in this region were stripped from orbiting satellite galaxies.

\vspace{-0.5cm}

\section*{Data Availability}
All data used in the analysis of this paper can be downloaded from the Mikulski Archive for Space Telescopes, https://archive.stsci.edu, or the ESO Science Archive Facility, http://archive.eso.org.

\vspace{-0.5cm}

\section*{Acknowledgements}
For the purpose of open access, the author has applied a creative commons attribution (CC BY) to any author accepted manuscript version arising.

We thank the reviewer, Chris Collins, for useful feedback and Leonardo Ferreira for helping with the technical details of {\it HST} data reduction. SW thanks the University of Nottingham, UK, Vice Chancellor International Scholarship. NAH thanks the Science and Technology Facilities Council, UK, consolidated grant ST/T000171/1. GN thanks the Natural Sciences and Engineering Research Council Canada, Discovery Grant and Discovery Accelerator Supplement and the Canadian Space Agency, grant 18JWST-GTO1. YMB thanks the Dutch Science Organisation (NWO), Veni grant 639.041.751. JM thanks the George P. and Cynthia Woods Mitchell Institute for Fundamental Physics and Astronomy at Texas A\&M University. DW thanks the German Research Foundation, Emmy Noether Grant WY 179/1-1, the Daimler and Benz Foundation and the German Space Agency, Verbundforschung grant 50 OR 2213.
Based on observations collected at the European Organisation for Astronomical Research in the Southern Hemisphere under ESO programmes 094.A-0343(C), 094.A-0343(D), 094.A-0343(E), 096.A-0317(A), 096.A-0317(B), 096.A-0317(C).




\vspace{-0.5cm}

\bibliographystyle{mnras}
\bibliography{Bibliography} 



\appendix

\label{lastpage}
\end{document}